# Near decomposability of complex networks based on Simon's theory of complexity architecture: conceptions, approaches and an application[i]

Jianmei Yang

School of Business Administration, South China University of Technology, Guangzhou 510640

**Abstract**

Based on Simon’s theory of complexity architecture, we investigate the near decomposability of complex networks, including network near decomposable hierarchy (NDH) and near decomposability in behavior and properties (NDBP): the related conceptions and approaches are given and illustrated with an example. Different from existing researches, we emphasize the network NDBP; and follow Simon's generation mechanism of complex things to find the network NDH; In addition, based on Simon's theory that most things in the real world are complex systems with the NDH; and the situation that the approach for analyzing the NDH of a simple room system in Simon's example requires that the system adjacency matrix be in the near block diagonal form, which is often not the case for the adjacency matrices of complex systems in the real world, we propose an approach for analyzing the NDH of complex networks (complex systems) by splitting network adjacency matrices, which is suitable for solving real-world problems faced rather than for scientific discovery. The near decomposability of complex networks, due to the advantages of network language, can provide a more convenient and effective way than the near decomposability of complex systems for understanding and dealing with the problems of complex things.

## 1. Introduction

Complex things may have various generation mechanisms. In the theory of complexity architecture [1-2], H. A. Simon, Nobel laureate in economics and one of the founders of Complexity Science, Artificial Intelligence and other disciplines, reveals one generation mechanism of complex things and pointes out that complex things often have a property based on the mechanism : near decomposability (see Section 2). The near decomposability of complex things can be divided into two parts: the nearly decomposable hierarchy (NDH) of the structure and the near decomposability in behavior and properties (NDBP). The NDBP is concerned with the characteristics and differences in the short- and long-run behavior and properties of complex things.

Complex networks [3-6] and complex systems are both models of complex things, while the former uses the language of networks and the latter uses the language of systems. Since complex things have the near decomposability, their models should also have the near decomposability. Simon has proposed the near decomposability of complex systems by modeling complex things with complex systems. In this paper, we investigate the near decomposability of complex networks (including the NDH and NDBP of networks).

The related conceptions and approaches are given and illustrated with an example of Input-Output complex network.

In the Network Era, investigating the near decomposability of complex networks is very necessary and important. This is because (1) network models are increasingly used than system models to analyze the problems of real-world complex things; (2) the near decomposability of complex networks, not only like the near decomposability of complex systems, can lead us to "see" the whole and the parts of a complex thing and thus understand it, but also can leads us to "see" the specific interactions between these parts to better understand it (see Section 5.1).

Although we have not seen the researches related to the NDBP of networks, there have been a number of researches related to the NDH of complex networks. Meunier et al. [7-8] based on functional magnetic resonance imaging measurements in 18 healthy volunteers, by using a multi-level method [9] find that human brain functional networks have a hierarchical modular organization; Ravasz et al. [10-11] use a hierarchical clustering algorithm [12] to analyze the metabolic networks of 43 distinct organisms and find that two features (the “scale-free” and “high degree of clustering) of many complex networks are the result of hierarchical organization; Clauset et al. [13] detect the hierarchical structure of networks by fiting the hierarchical model to observed network data by using the tools of statistical inference, combining a maximum-likelihood approach with a Monte Carlo sampling algorithm; Song et al. [14] use a box-covering technique to coarse-grain complex networks, resulting in the network hierarchy; Communities of complex networks are seen as groups of vertices within which connections are dense, but between which connections are sparser [15-16], so the community detection can give a nearly decomposable structure of a complex network. The methods for community detection are the basis of a number of methods for analyzing network hierarchies [7-8]. Some community detection methods are related to methods of hierarchical clustering [17], in which the GN algorithm [15] and the fast algorithm [18] identify communities based on a hierarchical tree composed of different sizes of communities.

Below, the possible scientific contributions of this paper are given in an overall comparison with the related researches mentioned above ( a one-to-one comparison with them is given in Section 5.2).

(1) Unlike related researches, we investigate the near decomposability of complex networks based on Simon's complexity architecture theory, that is, we not only focus on analyzing the network's NDBP in Simon’s sence, i.e., analyzing the change laws of the network's short- and long-run behaviors and properties, but also analyze the network's NDH in accordance with Simon's generation mechanism of complex things (see Section 2). This provides a new idea to the current research on hierarchical organization of complex networks.

(2) We develop a new approach for analyzing the NDH of complex networks. The novelty of this approach is that the adjacency matrix of a network is first split into several matrices of the same order as the adjacency matrix according to the order of magnitude (OOM) of vertex interactions, and then these split matrices are analyzed separately using methods of graph theory, and finally the results of the analyses are synthesized to obtain the NDH of the network; whereas the related approaches, by using quantitative methods based on graph theory, statistics and numerical simulations, etc., directly analyze the entire adjacency matrix to obtain the the NDH of the network. However, it is important to note that the meaning of OOM (which should be in quotes) here is not exactly the same as that in pure mathematics and the OOM differences refers to “qualitative” differences. The criteria for dividing the “qualitative” (OOM) differences of some property of a complex thing in real world are not usually determined mathematically, but rather qualitatively, based on the knowledge of the discipline to which the property belongs and the required solution accuracy of the problem faced, so the approach is more suitable for application-oriented disciplines (focus on solving the problems faced, not on scientific discoveries). We believe that a predominantly quantitative approach with some qualitative analysis is a good approach for application-oriented disciplines, as long as the qualitative analysis is empirically or academically based and the approach solves the problem satisfactorily. For example, using our approach, the example in this paper satisfactorily analyzes a chain price increasing (CPI), so it can be said that our approach is a good one.

(3) We extend Simon’s approach for analyzing the NDH of 2-level simple systems to complex systems (complex networks). First, the approach for analyzing the NDH of complex systems consists of two parts: the method of dividing the NDH levels and the method of finding subsystems. Simon has given an approach for the NDH analysis of a room system (see Section 3.3.1), which is essentially a method for finding subsystems. Since the adjacency matrix of the system (Fig. 2(a)) can be divided into 2 different clusters, which implies that the NDH of the system has only 2 levels, therefore after finding the subsystems by Simon's method, the NDH of the system is also found. But since a real complex system is made up of a large number of parts that have many interactions (see Section 2), the adjacency matrix is usually not of the block diagonal form, much less this form with two blocks, so Simon's approach is just a way to analyze the NDH of 2- level simple systems rather than complex systems, for complex systems it is only the method for finding subsystems. We split the adjacency matrix of a complex system according to the OOMs of the interactions into several matrices (matrix 1 , 2...N, see Section 3.3.1), each with its own near block diagonal form, which not only gives a method for dividing the NDH levels, but also creates the conditions for continuing to use Simon's method to find subsystems at each level, thus extending the approach of Simon's NDH analysis to complex systems with non-block diagonal adjacency matrices. Second, Simon did not detail how to compute the various interactions of an aggregative vertex from a subnetwork (AV) which are also explored in this paper.

(4) We reveal the law of CPI of industries of an actual Input-Output complex network and give a specific analysis approach for the NDH and NDBP of general Input-Output complex networks

## 2. Theoretical foundation

Simon's theory of complexity architecture is the theoretical foundation of this paper and the relevant contents are reviewed below.

*Simon’s generation mechanism and the near decomposability of complex things:* Simon argues that just like matter at the microscopic level (elementary particles, atoms, molecules, macromolecules), complex things are often generated in a hierarchical and aggregate manner from bottom to top, according to the order from strongest to weakest of the OOM of interactions; and this generation mechanism determines that complex things often have the near decomposability (including the NDH and NDBP of complex things).

*The near decomposability of complex systems:* This is a formulation of the near decomposability of complex things in the language of systems by Simon. The near decomposability of complex systems includes the NDH and NDBP of systems. By a complex system, Simon means one made up of a large number of parts that have many interactions; in such systems, given the properties of the parts and the laws of their interactions, it is not a trivial matter to infer the properties of the whole [2, p. 183].

The NDH of complex systems means (1) A complex system is hierarchical: the system is composed of interrelated subsystems, each of latter being in turn hierarchic in structure until we reach some lowest level of elementary subsystem [2, p. 184]. (2) A complex system is nearly decomposable: the interactions among subsystems at different levels are of different OOMs and the further down the hierarchy, the stronger the interactions. In other words, at each level, the interactions between subsystems are weaker to negligible than the interactions within each subsystem, but still need to be embodied in an aggregate way in the upper levels. Simon further tells us by a watchmaker's parable [2, p. 188] that complex systems often have the NDH because it is the NDH among the possible complex forms that allow them to have the time needed for evolution. Fig. 1(a) shows the NDH of a room system (although the room system is not made up of a large number of parts, it also has the near decomposability) to illustrate the meaning of the NDH of complex systems.

Simon also states the meaning of the NDBP of systems through the heat exchange behavior of the room system [2, p. 198]: (1) the short-run behavior ("behavior" used by Simon includes properties) of each of the component subsystems (e.g. A, B and C in Fig. 1(a)) is approximately independent of the short-run behavior of the other components. (2) In the long run, the behavior of any one of the components depends in only an aggregate way on the behavior of the other components.

*Epistemological significance of complexity architecture theory:* First, the near decomposability of complex things in this theory gives us confidence to understand complex things. This is because it lets us know that although the NDH makes them fill the world; the NDBP makes most of them only weakly connected to each other, so we only need to find a tiny fraction of all possible interactions of a complex thing with other things to understand it. Second, the near decomposability of complex systems in this theory, using the language of systems, provides us with an approach to find and consider the tiny fraction.

## 3. Near decomposability of complex networks: a network formulation of the near decomposability of complex things

### 3.1. Constructing complex network models of complex things and features of the models

Constructing: first, it needs to be noted that a specific complex thing often has several types of interactions, so it often has more than one complex network model. However, when faced with a real-world problem, only one network model for this complex thing needs to be constructed based on only one type of interactions relevant to the problem.

The elementary parts of a complex thing are regarded as vertices; if there is some type of interactions that we are concerned about between two elementary parts, an edge is connected between the corresponding vertices; and the intensity of such interaction is used as the edge weight. In this way, a weighted complex network model of this complex thing for the problem is obtained. The interactions may be directed or undirected, so the corresponding network model (short for network) may be directed or undirected. And if the network vertices are represented by the sequence numbers of their vertex sets, and $a_{ij}$ denotes the weight of the directed edge from vertex i to j, then the elements of the adjacency matrix of the network are $a_{ij}$ ( i, j=1, 2…n). Imitating this definition, the room network of the room building is shown in Fig. 1(b).

Features: the complex network models of complex things have three features. First, such networks are formed by the interaction relationship, excluding other relationships. Of course, the networks formed by other interactions (e.g. similarity relationships) may also be transformed into networks of interactions with some additional definitions. Second, such networks must be weighted networks, since the near decomposability of complex networks is based on the intensity of vertex interactions. Third, the undirected graph corresponding to such a network must be connected, because its any vertex represents an elementary part of a complex thing that must interact with at least one other elementary part of the thing to enter the thing, i.e., there must be a path between any vertex of the network and its other vertices.

### 3.2. Conceptions of the near decomposability of complex networks

We propose following conceptions of the NDH and NDBP of complex networks. Because "subnetworks" replace "subsystems", these conceptions of complex networks are very different from those conceptions of complex systems.

#### 3.2.1. Conception of the NDH of complex networks

The NDH of complex networks has two meanings. (1) A complex network is composed of some subnetworks, each of the latter is in turn hierarchic in structure until we reach some lowest level of elementary subnetwork. Here, according to Simon's generation mechanism, a subnetwork at any level is defined as a connected component consisting of some single vertices (original vertices), or some single vertices and AVs of the network, which are linked by the interactions having the OOM corresponding to this level. The interactions among vertices within any subnetwork at each level are much weaker than those within any subnetwork at its next level. The interactions among vertices within any subnetwork at different levels often are of different OOMs. The lower the level of a subnetwork is, the stronger and earlier the interactions among its vertices are. Note: the NDH of a directed- weighted complex network has two types of displays: NDH not showing directions and weights and NDH showing directions and weights. The former only shows the presence or absence of interactions between vertices, and the subnetworks at all levels are the simplest networks: no direction, no edge weights and no loop (loop: denoting self-to-self interaction). But the two displays essentially correspond to the same NDH, with only the difference between showing and not showing the directions and weights. While for undirected-weighted complex networks there are two displays of the NDH: showing and not showing the weights (2) The actual generation process of the NDH of a complex network is from bottom to top: The single vertices of the network having strongest interactions (OOM 1) first compose some level-1 (lowest-level) subnetworks;

and then, some of level-1 subnetworks, each as an AV, are linked to each other or to a number of other single vertices by the weaker interactions (OOM 2) to compose some level-2 subnetworks, some other level-1 subnetworks as AVs may enter higher-level subnetworks; This process continues until one highest-level subnetwork is composed by the weakest interactions (OOM N). The emergence of one highest-level subnetwork means that all single vertices of the network will directly or indirectly enter final one AV. Note: according to Simon's generation mechanism, the stronger the interaction the earlier it occurs, which higher level an AV formed by a lower-level subnetwork will enter depends on the corresponding OOM of the maximum value of the aggregative interactions of this subnetwork to all external vertices; while the aggregative interaction of one subnetwork (i.e., the interaction of its AV) to an external vertex is defined as the maximum value of the one-to-one interactions of all internal vertices of this subnetwork to this external vertex.

The NDH of the room network (not showing weights) is shown in Fig. 1(c). It is obviously different from the NDH of the room system (Fig. 1(a)). Although they both have three levels, the former can be decomposed into one level-2 and three level-1 subnetwork while the latter into three level-2 and eight level-1 subsystems. Fig. 1(d) shows the generation process of the room network. First, the vertices $A_1$, $A_2$ and $A_3$; $B_1$ and $B_2$; $C_1$, $C_2$ and $C_3$ are respectively linked by the strong heat diffusion to compose three level-1 subnetworks. Then, these level-1 subnetworks each as a AV (A, B and C) are linked by the weak heat diffusion to compose one level-2 subnetwork, which means that $A_1$, $A_2$ and $A_3$; $B_1$ and $B_2$; $C_1$, $C_2$ and $C_3$ will all indirectly enter AV N.

**Fig. 1.** Various schematics of the room building

### 3.2.2. Conception of the NDBP of complex networks

The NDBP of complex networks includes the near decomposability in behavior and properties of networks. *The near decomposability in behavior of networks:* For each level of the NDH, in the shorter run, all subnetworks are approximately independent of each other in terms of the interaction OOM at that level, but there are interactions having this OOM among the vertices within any subnetwork; in the longer run, the interaction between two vertices belonging to different subnetworks at that level are embodied in an aggregate way by the interaction between their respective corresponding AVs at some higher level. Take the room network as an example. In the short run, its three level-1 subnetworks are approximately independent of each other in terms of the OOM-1 interactions, the OOM-1 interactions only occur among the vertices within each level-1 subnetwork. In the long run, the OOM-2 interactions occur among the vertices A, B, and C of the level-2 subnetwork. A, B and C are three AVs formed respectively by one level-1 subnetwork. The reason for A, B and C having OOM-2 interactions is that there are originally weak OOM-2 interactions between the vertices of different level-1 subnetworks, such as between $A_1$ and $B_{1,}$ however, due to the strong OOM-1 interactions make $A_1$, $A_2$ and $A_3$; $B_1$ and $B_2$; $C_1$, $C_2$ and $C_3$ form rapidly an AV (A, B and C) respectively, the original weak OOM-2 interactions between those vertices of different level-1 subnetworks can only be embodied by the interactions among their AVs.

*The near decomposability in properties of networks*: We argue that from the near decomposability in a certain behavior, the near decomposability in the property directly related to that behavior can be derived: In the shorter run, the properties of the vertices within each level-1 subnetwork will first reach an equilibrium state, independent of the other level-1 subnetworks and vertices; In the longer run, the properties of the vertices within each level-2 subnetwork will reach an equilibrium state in the same way, and so on to the higher levels. Here, the property of an AV of a higher-level subnetwork is the aggregative property of all vertices of the lower-level subnetwork which form the AV. For example, from the near decomposability in heat diffusion behavior, the near decomposability in the temperature of the room network can be derived: In the short run, the vertices within each level-1 subnetwork, such as $A_1$, $A_2$ and $A_3$, will reach a state of thermal equilibrium, independent of other level-1 subnetworks and vertices. In the long run, when there is an almost uniform temperature throughout the room network, the AVs A, B and C within the level- 2 subnetwork will reach this equilibrium state.

### 3.3. Approaches for analyzing the near decomposability of complex networks

#### 3.3.1. Approach for analyzing the NDH of complex networks

First we review the approach for analyzing the NDH of complex systems. We already know that the approach for analyzing the system NDH consists of two parts, and for complex systems Simon's approach for analyzing the room system only details how to find subsystems. We introduced Simon's theory about the architecture of complexity to the study of industry systems [19]. In ref. [19], according to Simon's theory and with reference to the practice in the economics community we summarized a division method of the hierarchy levels of industry systems and then proposed an approach for analyzing the NDH of industry systems by splitting a system adjacency matrix which does not have the near block diagonal form [19]. This work of ours was affirmed by Simon (see Supplementary Material I), and referred by scholars to as a reducing space approach for dealing with complexity [20].

In this paper, we develop the idea of splitting matrices to complex networks, and according to the conception of NDH of complex networks in Section 3.2.1 proposes the following approach (including logical steps and general methods) for analyzing the NDH of complex networks by splitting network adjacency matrices. Note that this approach is better understood in combination with the NDH analysis process of the room network in Fig. 2. The reason why the process of the room network is used as an example is that the network is small enough to show the data of the analysis process directly, thus providing a clearer illustration of the approach. However, more details of this approach need to be shown with the example of the input-output complex network in Section 4.

(1) Splitting the adjacency matrix of a network. The adjacency matrix of the complex network model of a real-world problem often does not have a near block diagonal form, and we need to split the adjacency matrix of the network. First, According to the knowledge of the discipline to which the problem faced belongs and the required solution accuracy, the elements of the adjacency matrix are arranged into N categories in order from largest to smallest: category 1, 2… N. Category 1, 2… N are intensity criteria of the vertex interactions of OOM 1, 2…N, respectively. Note: There are N OOMs of vertex interaction intensity, which determine that the network can be decomposed into N levels; The OOM 1, 2…N represent the interaction intensity of vertices within each subnetwork at level 1, 2…N (from bottom to top), respectively. Moreover, the determining the criteria is a trial-and-error process: the number of the categories and the range of element values for each category should be tested several times until the resulting NDH provides satisfactory insight into the problem faced. Second, according to above N categories of elements, the network adjacency matrix is split into N matrices of the same order: matrix 1, 2...N. The element values of matrix 1 (including diagonal elements) either are in the range of category 1 or are 0, the element values of matrix 2 are in the range of category 2 or are 0, etc. Note: matrix 1, and matrix 2...N (after aggregations, see Section (3)), which show the the presence or absence, direction and intensity of vertex interactions of OOM 1, 2 … N, respectively, all have block diagonal forms.

(2) Writing the interaction matrices of OOM 1, 2…N. The non-zero elements in matrix 1, 2...N are changed to 1 and the Boolean matrices at each OOM of interactions will be obtained. These Boolean matrices are called the interaction matrices of OOM 1, 2…N, respectively. The interaction matries of OOM 1, 2…N differ from matrix 1, 2...N in that they do not show the intensities of interactions and are 0-1 matrices.

(3) Finding subnetworks at each level of the NDH not showing directions and weights. First, to clarify a few points: according to the conception of the NDH, a subnetwork at any level is a connected component whose vertices are linked by the interactions having the OOM at that level (excluding self-to-self interactions), two vertices are connected regardless of who has the interaction OOM at that level on whom, so the connected components of a directed network are the connected components obtained by regarding it as an undirected network. We call the matrices of the undirected networks corresponding to the interaction matrices of OOM 1, 2…N their respective symmetric matrices. These symmetric matrices differ from the interaction matrices in that they do not show the directions of interactions and are symmetric matrices. The subnetworks at each level are obtained by calculating the connected components of the symmetric matrix of the interaction matrix (or the interaction matrix after aggregations, see below) corresponding to that level and the number of subnetworks is equal to the number of calculated connected components. Second, the specific finding process is as follows: The level-1 subnetworks are obtained by calculating the connected components of the symmetric matrix of the OOM-1 interaction matrix. According to the conception of the near decomposability in behavior of networks (see Section 3.2.2), the OOM-2 interactions between the vertices which belong to different level-1 subnetworks are embodied in an aggregate way by the interactions between their AVs. Therefore, different from the finding level-1 subnetworks, before finding level-2 subnetworks, in the symmetric matrices of interaction matrices from OOM 2 to OOM N, the name of each level-1 subnetwork (which is also the name of the AV formed by it, the same below), should be used respectively to replace the vertex sequence numbers contained by this subnetwork. In this way, we get the symmetric matrices of the interaction matrices from OOM 2 to OOM N after first aggregation. Then, by finding the connected components of the symmetric matrix of the OOM-2 interaction matrix after first aggregation, all level-2 subnetworks are obtained. Similarly, before finding level-3 subnetworks, first, in the symmetric matrices of OOM-3 to OOM-N interaction matrices after first aggregation, the name of each level-2 subnetwork should be used respectively to replace the vertex sequence numbers and the names of the AVs which are contained by this level-2 subnetwork; and the symmetric matrices of OOM-3 to OOM-N interaction matrices after second aggregation are obtained. Then by finding the connected components of the the symmetric matrix of OOM-3 interaction matrix after second aggregation, the level-3 subnetworks are obtained, and so on. Based on the symmetric matrix of OOM-N interaction matrix after the (N-1)th aggregation, one level-N subnetwork will be found and all single vertices are contained directly or indirectly by the subnetwork.

(4) By arranging these subnetworks from 1level 1 to level N in the order from the lowest to the highest level, the NDH not showing directions and weights of the complex network is obtained.

(5) For a directed-weighted complex network, its NDH showing directions and weights can be obtained by labeling the directions and weights of edges to the above NDH not showing directions and weights, the data of these directions and weights can be taken from the matrix 1, 2,...N in step (1). In addition, the NDH showing directions and weights also can be obtained by directly calculating the connected components of the matrix 1,and matrix 2...N after aggregations; moreover, in this way, the NDH expressed in other form subnetworks, such as subnetworks with loops, also can be obtained.

**Fig. 2**. The process of analyzing the room network NDH

### 3.3.2. Approach for analyzing the NDBP of complex networks

After obtaining the NDH of a complex network, it is possible to analyze the NDBP of the network. However, unlike the NDH, there are no common logical steps to analyze the NDBP. And in general methods, there are the following common points. (1) If the network is a directed-weighted network, in order to analyze its NDBP we should use the subnetworks of its NDH showing directions and weights. (2) The interaction of an AV to an external vertex is defined as the maximum of the one-to-one interaction of all vertices within the subnetwork forming this AV to this external vertex. Moreover, the aggregative properties of these internal vertices are the properties of this AV. Other specific methods for analyzing the NDBP depend on the discipline knowledge to which the behavior and properties belong. For example, when analyzing the near decomposability in heat diffusion behavior and temperature of the room network, the methods of heat diffusion coefficients of thermodynamics and temperature measurement should be adopted; whereas to analyze the near decomposability in input behavior and prices of an Input-Output network, the methods of direct consumption coefficients [21] and price transmission of Input-Output economics should be adopted (see Section 4.3).

## 4. An application

The industry sectors of Guangdong province (GD for short) is a complex thing, and we focus on its Input-Output interactions in the year *X*. Based on GD Input-Output direct consumption coefficient table of 139 sectors (industries), we first construct an Input-Output complex network model of the complex thing, and then analyze the NDH and NDBP of the complex network.

### 4.1. Constructing the Input-Output complex network model of GD industry sectors

The Input-Output complex network model of GD industry sectors, referred to as GDIO network, is shown schematically in Fig. 4(a). The network is a connected and directed network with double weights and loops. In the network, there are 139 vertices (each vertex is called a single-industry vertex) representing 139 industries. A directed edge $e_{ij}$ of the network indicates there is an Input-Output interaction between Industries i and j, and the arrow points to the inputted Industry j. The Input-Output direct consumption coefficient $a_{ij}$ (i, j= 1, 2 …139) is chosen as the weight of edge $e_{ij}$, i.e., the input intensity of Industry i to j. In this way, the direct consumption coefficient matrix of 139 industries is the adjacency matrix of the network. In the adjacency matrix, the vertex representing certain industry is denoted by the sequence number of the industry in the Input-Output table. The rationale for choosing the direct consumption coefficients as the weights of the directed edges of the network is that, according to Input-Output economics, the direct consumption coefficient $a_{ij}$ represents the input quantity of Industry i to Industry j for one unit output of Industry j, so the $a_{ij}$ is relatively ideal means that measures the input intensity of Industry i to Industry j.

### 4.2. The NDH of GDIO network

*Process of analysis:* First, splitting the adjacency matrix of GDIO network. After several trials, finally the direct consumption coefficients $a_{ij}$ are arranged into five categories. $1 > a_{ij} \geq 0.3$ are of category 1; $0.3 > a_{ij} \geq 0.19$ — category 2; $0.19 > a_{ij} \geq 0.1$ — category 3; $0.1 > a_{ij} \geq 0.06$ — category 4; $0.06 > a_{ij} > 0$ — category 5. Categories 1 to 5 are input intensity criteria of OOM 1 to 5, respectively. Thus, the adjacency matrix of the network is split into 5 matrices of the same order. The network can be decomposed into 5 levels and OOM 1 to 5 are the input OOMs at level 1 to level 5 respectively. Next, follow the remaining steps in Section 3.3.1 to find the subnetworks at each level to obtain the NDH of GDIO network. We use Gephi software (https://gephi.org/) to calculate the connected components [22] of the input matrices. Before using this software to enter network edge data, according to the meaning of the subnetwork, we removed the "loops", and only the one with the largest edge weight is retained when there are double edges between a pair of vertices.

*Results of analysis:* GDIO network is a directed-weighted network, so its NDH has two types of displays. For the clarity, only the NDH not showing directions and weights is shown (Fig. 3). For comparison, we show in Appendix I the NDH of GD industry system.

Because the input has five OOMs, the network can be decomposed five levels. First, from top to bottom, at level 5, GDIO network is composed of one level-5 subnetwork “5sn” (purple), it is a complete graph having five vertices. At level 4, subnetwork 5sn is decomposed into one level-4 subnetwork “4sn” (orange) and four single-industry vertices which are isolated vertices because they do not have OOM-4 input with any vertex. At level 3, subnetwork 4sn is decomposed into seven level-3 subnetworks from “3sn1” to “3sn7” (blue), one level-1 subnetwork “1sn9” (red) and several single-industry vertices. At level 2, subnetwork 3sn1 is decomposed into seven level-2 subnetworks “2sn1”, “2sn2”, etc (brown), three level-1 subnetworks “1sn5”, “1sn8”, “1sn10” (red) and several single-industry vertices; subnetworks 3sn2 to 3sn5 are decomposed into one level-2 subnetwork each and several single-industry vertices; subnetworks 3sn6 and 3sn7 are decomposed into two single-industry vertices respectively. At level 1, 11 level 2 subnetworks are decomposed into 11 level-1 subnetworks (excluding 1sn9) and a number of single-industry vertices.

Note that the sum of the numbers of single-industry vertices at level 4, 3, 2 and the number of single-industry vertices of all level-1 subnetworks is equal to the number of total industries —139.

Second, from bottom to top, as the result of OOM-1 input, some industry vertices of 139 industries compose respectively 12 level-1 subnetworks. As the result of OOM-2 input, the eight level-1 subnetworks, each as one AV and some single-industry vertices compose respectively six level-2 subnetworks; the other five Level-2 subnetworks are each composed by a number of single-industry vertices. As the result of OOM-3 input, all the level-2 subnetworks, three level-1 subnetworks, each as one AV, and some single-industry vertices compose a total of five level-3 subnetworks; other two level-3 subnetworks are each composed by a number of single-industry vertices. As the result of OOM-4 input, all level-3 subnetworks and level-1 subnetwork 1sn9, each as one AV, and some single-industry vertices compose one level-4 subnetwork 4sn. Finally, as the result of OOM-5 input, level-4 subnetwork 4sn, as one AV, and four single-industry vertices compose one level-5 subnetwork 5sn, which means that all the single-industry vertices of the network will directly or indirectly enter final one AV. In the generation process of GDIO network, the sum of the numbers of single-industry vertices sucked into the subnetworks at level 1 to 5 is also equal to the number of total industries —139.

**Fig. 3.** The NDH (not showing directions and weights) of GDIO network

### 4.3. The NDBP of GDIO network

Since GDIO network is a directed-weighted network, we use its NDH showing directions and weights to analyze its NDBP, where behavior and properties refer to the input behavior and prices of the industries.

#### 4.3.1. The near decomposability in input behavior of GDIO network

In terms of the input OOM at each level, the subnetworks at that level are independent of each other. That is, the input with the OOM corresponding to each level occurs only among the vertices (industries) within every subnetwork at that level: the OOM-1 (strongest) input occurs only among the vertices within each of the 12 level-1 subnetworks, the OOM-2 (stronger) input occurs only among the vertices within each of the 11 level-2 subnetworks and so on. Here, the vertices of a subnetwork at level 2 and above often contain one or more AVs. It is easy to derive that the input behavior of other Input-Output networks also has the near decomposability in input behavior similar to that of GDIO network.

#### 4.3.2. The near decomposability in prices of GDIO network

The industry prices are the property directly related to their input behavior, so the near decomposability in the input behavior will lead to the near decomposability in prices of GDIO network: If some reason causes price fluctuations in the industry vertices of the network, first, in the shorter run, the vertex prices within each of the 12 level-1 subnetworks will reach an equilibrium state, independent of the other level-1 subnetworks and vertices; then, in the longer run, the vertex prices within each of the 11 level-2 subnetworks will reach an equilibrium state in the same way, and so on up to the highest level.

### 4.3.3. The CPI in GDIO network

Other conditions being equal, the price increase of an industry has the knock-on effects on the prices of the relevant industries due to the increases of the costs. The process and characteristics of the CPI in an industry network is the reflection of the near decomposability in prices of the network. The following is a theoretical analysis of the CPI triggered by industry 1 (agricultural products) in GDIO network (Fig. 4(a)) based on Section 4.3.2.

(1) Methods of analysis: *Round-by-round calculation*. The CPI within any subnetwork of GDIO network is calculated round by round. Taking a level-1 subnetwork as an example, in round 1, we compute only the price increases of the vertices directly inputted by the original-price-increase industry; in round 2, we compute only the price increases of the vertices directly inputted by the industries whose prices have increased in round 1, and so on, until the CPI in this subnetwork is over. *Equilibrium condition:* Assume that an industry vertex is considered to be in a state of price equilibrium (no more price increase) if its single-round price increase is less than 0.1%; and when every industry vertex of a subnetwork is in this state, the subnetwork can be said in the state of price equilibrium. *Aggregative interaction:* The maximum value of the direct consumption coefficients of an external vertex to all vertices within a subnetwork is taken as the aggregative interaction of the subnetwork to that external vertex, which is the input coefficient of the AV formed by the subnetwork to that external vertex. *Aggregative property*: The arithmetic average of the price increases of all vertices of a subnetwork is taken as the aggregative property of the subnetwork, which is the price increase of the AV. *Calculation of price change:* According to the economics literature [23], if only the price of Industry i changes by $\Delta p_i$ (%), $a_{ij}$ is the direct consumption coefficient of Industry j to i, then the price of Industry j will show a cost-push change, and the price change of Industry j directly caused by the price change of Industry i is $\Delta p_j$ (%) = $\Delta p_i$ (%)* $a_{ij}$.

(2) Process of analysis: If the price of Industry 1 increases by 10%, this will first (implying a short time interval from the price increase of industry 1, i.e., a short run) affect the other six industry vertices within the level-1 subnetwork 1sn2 where it is located, due to the OOM-1 input to them of Industry 1 (Fig. 4(b)). We can calculate that among the six industry vertices, the $1^{st}$- round price increases of Industries 12, 13, 14, 15, 3 and 16 are 4.61%, 4.82%, 3.85%, 3.06%, 2.06% and 0.95%, respectively. Since Industry 14 has the internal input with a direct consumption coefficient of 0.3467 (Industry 14 has no input to other industries), its $1^{st}$-round price increase of 3.85% will in turn cause its own $2^{nd}$-round price increase of 1.34% (3.85%*0.3467), and its $2^{nd}$-round price increase will in turn cause its own $3^{rd}$-round price increase of 0.46% ($3.85\%*0.3467^2$), and so on. The common ratio 0.3467 is less than 1, the geometrical sequence of single-round price increases of Industry 14 will converge to 0, and after several rounds the price increase of Industry 14 will be very small. When the equilibrium is reached (it takes some time, implying a longer time interval from the price increase of industry 1, i.e., longer run), we can calculate that in the CPI, the price increase of Industry 14 is 5.81% of the sum of the first four rounds (its $5^{th}$-round increase is 0.056%< 0.1%), while other industries still maintain own $1^{st}$-round price increases; and the arithmetic average of the increases of the seven vertices of subnetwork 1sn2 is 4.47%. The 4.47% as the aggregative property of subnetwork 1sn2 is the price increase of AV 1sn2 (formed by subnetwork 1sn2) of level-2 subnetwork 2sn2. Next, the 4.47% price increase of AV 1sn2 will spread to other vertices within subnetwork 2sn2, due to its OOM-2 input to these vertices. From Fig. 4 (c), we can see that AV 1sn2 not only has OOM-2 input to other vertices within subnetwork 2sn2, but also has an internal input with a direct consumption coefficient 0.2839 (caused by Industry 15 aggregated by AV 1sn2). Therefore, after the 1st-round price increases of vertices 5, 28, 127, AVs 1sn6 and 1sn2 caused by the initial increase 4.47% of AV 1sn2, the 1st-round increase of AV 1sn2 itself will cause the 2nd-round price increase of other vertices and itself. Similarly, the 2nd-round increase of AV 1sn2 will continue to cause their 3rd-round price increase, and so on. When all industry vertices within subnetwork 2sn2 are in equilibrium, we calculate that the arithmetic average of the increases of these industries of subnetwork 2sn2 is 2.51%. The 2.51% is the price increase of AV 2sn2 of level-3 subnetwork 3sn1 (Fig. 4 (d)). Again, due to the OOM-3 input, the 2.51% price increase of AV 2sn2 will spread to some vertices within the subnetwork 3sn1. The same method can be used to calculate that the price increase of AV 3sn1 of level-4 subnetwork 4sn (Fig. 4 (e)) is 0.124%. We know that the direct consumption coefficients corresponding to the OOM-4 inputs are all less than 0.1, so within the subnetwork 4sn, the price increases of other vertices caused by the 0.124% increase of AV 3sn1 are much smaller than 0.1%, and this mean the CPI is over before it reaches subnetwork 4sn, far from reaching subnetwork 5sn (Fig. 4 (f)). So far, we know that from bottom to top, through the interactions of OOMs 1, 2, and 3, which industries are affected directly or indirectly (through AV) by the 10% price increase of industry 1, and the magnitude of their respective price increases.

(3) Characteristics: From the above price increase process triggered by Industry 1, we can draw the following characteristics of the CPI. *Minority*: the price increase of Industry 1 only affects 18 single-industry vertices (including Industry 1), while this industry network has 139 single-industry vertices. *Stage by stage:* Temporally, the CPI occurs in stages and sequentially within a relevant subnetwork at level 1 to level 3. *Batch by batch:* from a “spatial” perspective, the CPI in each stage affects a group of vertices within a subnetwork associated with that stage, so the CPI occurs in batches. *Aggregate way*: Except for the CPI in the level-1 subnetwork in stage 1, which is directly triggered by the original price increase single-industry, the CPI in the level-2 or higher subnetwork in other stages is caused by one AV associated with that stage, respectively. Moreover, the input of each AV to any industry is the aggregative input of its corresponding subnetwork, and the magnitude of AV price increases reflects the aggregative property of its corresponding subnetwork.

Note: For GDIO network, the block diagonal form of the symmetric matrix of OOM-1 interaction matrix (139*139) is shown in Fig. (1) o Appendix II, from which it can be seen that 41 out of the 139 industry vertices are divided into 12 (OOM-1) blocks; the block diagonal forms of the OOM-1 interaction matrix itself and matrix 1 (see Section 3.3.1) are also given (Fig. (2), (3) in Appendix II); moreover, the symmetric matrix of OOM-2 interaction matrix after aggregation can be divided into 11 OOM-2 blocks involving 8 level-1 subnetworks, each as one AV, with 24 single-industry vertices (blocks 1 to 11: 1sn 1, 81, 85, 87, 92, 93; 1sn2, 5, 28, 127, 1sn6; 1sn3, 44, 49; 1sn7, 1sn12, 110; 36, 37, 133; 2, 34; 9, 1sn4; 26, 29; 54,102; 116, 118; 1sn11, 95), etc..

**Fig. 4.** GDIO network and the CPI triggered by Industry 1 in this network

## 5. Discussion

### 5.1. Comparison of the near decomposability of complex networks and complex systems based on the example

We know that Simon discovers the near decomposability of complex things and has revealed it by the near decomposability of complex systems. So, what are the advantages of the near decomposability of complex networks such that it needs to be studied? First, since the network language can indicate the presence or absence, direction and intensity of one-to-one interactions between parts of complex things, while the system language cannot, the NDH of complex networks gives us a more detailed and deeper but also more intuitive and convenient understanding of the structure of complex things than the NDH of complex systems. Second, when dealing with one specific problem of a complex thing, the NDBP of either its complex system or complex network tells us that it is often necessary to focus on the behavior and properties of only one level-1 subsystem or subnetwork in the shorter run, and only in the longer run, the behavior and properties of one related level-2 subsystem or subnetwork, etc. However the NDBP of its complex system cannot point out directly which interactions need to be focused on, but can only tell which parts of the complex thing these interactions are in; whereas the NDBP of its complex network can directly point out what the specific interactions to be focused on are due to the network language, which allows us to understand the behavior and properties of complex things more quickly, and thus act more effectively to deal with specific problems.

Take GDIO network as an example. The NDH of GDIO network (Fig. 3) not only shows that GD industry sectors as a complex thing can be decomposed into 5 levels, which subnetworks each level has, the industry vertices (including the AVs) contained in each subnetwork and the input interactions and intensities among these industry vertices (showed by the NDH showing directions and weights); while the NDH of GD industry system (see Appendix I), on the other hand, shows that it has 5 levels, how many subsystems are in each level, and which industries are within each subsystem, but does not have any information on the input interactions and intensities within these subsystems. Moreover, the NDBP of GDIO network states: in the shorter run, when the price of industry 1 has just increased, just based on Fig. 4(b) to analyze the six weighted directed edges and one loop within level-1 subnetwork 1sn2 that have the strongest direct or indirect input interactions with industry 1, we can obtain the industries most affected by the price increase of industry 1 and the magnitude of their price increases. In the longer run, just based on Fig. 4(c) to analyze the four weighted directed edges and one loop in the level-2 subnetwork 2sn2 that have the next strongest input interactions with AV 1sn2, we can get the industries that are next most affected by the price increase of industry 1 and the magnitude of the respective price increases of these industries, etc. However, this is not possible through the subsystems 1ss2 (corresponding to 1sn2) and 2ss2 (corresponding to 2sn2) of GD industry system (see Appendix I) because there is no information about the interactions in 1ss2 or 2ss2.

### 5.2. One-to-one comparison of our NDH with related researches of complex networks

The purpose of the one-to-one comparison is mainly to show how our NDH differs from related researches of complex networks.

(1) The module-based complex network hierarchy of Meunier et al. or Ravasz et al. This is the most relevant research to our NDH. First, we study the hierarchy of general complex networks, Meunier et al. [7-8] and Ravasz et al. [10-11] study the hierarchy of specific complex networks. Our hierarchy consists of nested subnetworks, whereas the brain network hierarchy of Meunier et al. consists of nested submodules and the metabolic network hierarchy of Ravasz et al. consists of nested topologic modules. The building blocks of our hierarchy are subnetworks, those of Meunier et al. are submodules, and those of Ravasz et al. are topologic modules; and the subnetworks are formed by vertices that are reachable at the same OOM of interactions, the submodules are often made up of anatomically neighboring and/or functionally related cortical regions, and the topologic modules are composed of cellular components that are spatially or chemically isolated. From the above, we can see that the "complex network", "hierarchy" and "building blocks" in our NDH all have a much broader meaning. Second, our hierarchy is generated strictly according to Simon's generation mechanism. Meunier et al. state that their hierarchy analysis is related to the near decomposability of complex systems, but do not give a specific connection to Simon's generation mechanism. And Ravasz et al. state that the biological network with hierarchical topology may emerge by copying and reusing existing modules or motifs. Nevertheless, we think the approaches they actually use to find modules and the hierarchy show that their hierarchy analysis follow Simon's generation mechanism (i.e., based on the OOM of interactions and in an aggregate way), because for biological organizations, the intensity of interactions is reflected in spatially distances or chemical reactions [2, p. 187); and modules also imply aggregate way. Finally, to obtain the the NDH of a network, our approach requires first splitting the adjacency matrix of the network and then analyzing these split matrices separately with methods of graph theory; whereas Meunier et al. ( with the multi-level method) and Ravasz et al. (with the hierarchical clustering algorithm) directly analyze the entire adjacency matrix.

(2) The research of Clauset et al. The approach proposed by Clauset et al. to detect network hierarchy is based on statistics and computer simulation, while our approach is based on Simon's generation mechanism and graph theory**.**

(3) Coarse-graining of complex networks of Song et.al. In the process of coarse-graining of a complex network by the box-covering technique, at each level, a complex network is tiled with a minimum number of boxes, and finally the coarse-grained vertices of each level and a hierarchy of the complex network is obtained. Since the size of the box used at every level is constant implying the OOM of the interactions within a coarse grain at different levels is the same, the concept of the coarse grain is different from that of our subnetwork and the hierarchy of Song et al. is different from our NDH.

(4) Community detection: First, the conception of communities and subnetworks are similar, but communities are only similar to the level-1 subnetworks because the vertices of each community contain only the original vertices of a network, and no Avs. Second, the community structure is also a complex network nearly decomposable structure, but, unlike our NDH, this structure has only one level. Third, weighted or unweighted complex networks formed by any relationship have the communities, while weighted complex networks formed only by interaction relationship have the subnetworks. Fourth, from the GN algorithm and the fast algorithm, we can know the hierarchical tree in these algorithms is also a network hierarchy, but its each level corresponds to all original vertices, only different levels correspond to different community divisions for those vertices, so it is different from our NDH, and just a tool to perform the network community detection.

### 5.3. Limitations and further work

Limitations: (1) Our approach for analyzing the NDH of networks by splitting network adjacency matrices is somewhat subjective because it requires qualitatively determining the criteria for dividing OOMs of interactions. However, since this "qualitative determination" is based on the knowledge of the discipline to which the problem faced belongs, and since the determined criteria must also be tested and improved, according to the satisfactory degree of problem solving, it remains essentially objective. (2) Our approach for analyzing the NDH is used to solve problems, and therefore it is more suitable for application-oriented rather than scientific discovery disciplines. (3) When analyzing the near decomposability in properties of complex networks for real-world problems, the calculation results are approximate. For example, in our research of the CPI in GDIO network, the calculated price increases of affected industries are approximate, although the degree of approximation is high and the calculation process is very simple.

Further work: (1) We intend to do an empirical research on the near decomposability in prices of GDIO network. The brief idea is: First, the dynamic data which include the timing and magnitude of price increases for each industry in a real CPI triggered by one original-price-increase industry will be collected. Second, the collected data will be statistically analyzed to see whether the actual CPI is characterized by the Minority, Stage by stage, Batch by batch and In aggregate way. Third, the actual and theoretical data of the CPI are compared in order to improve our proposed analysis methods, focusing on the methods of determining the criteria of input OOMs and the calculation of price increase of AVs. (2) We will to do more example analysis of other real-world problems. (3) We plan to explore the connection between the generation mechanism of the hierarchical topology of Ravasz et al. and Simon's mechanism, as well as the connection between our approach for analyzing NDH and that of Meunier et al. and Ravasz et al.

**Acknowledgements**

The work was supported by the National Natural Science Foundation of China (Project No.71273093). The author thanks Xiangrong Liu for providing and Jingyi Zhang for pre-processing the Input-Output data used in the example; and Ruiqiu Ou and Quan Chen for discussions about economics.

## References

[1] H.A.Simon, The Architecture of Complexity, Proceedings of the American Philosophical Society 106 (6) (1962) 467–482.
[2] H.A. Simon, The architecture of complexity: hierarchic systems, The Science of the Artificia1 ,The MIT Press, Cambridge, MA (1996) 183-216.
[3] D.J.Watts, S.H.Strogatz, Collective dynamics of 'small-world' networks, Nature 393 (1998) 440-442.
[4] D.J.Watts, Small Worlds: The Dynamics of Networks between Order and Randomness, Princeton University Press, Princeton, NJ, 1999.
[5] A.-L. Barab ́asi, R. Albert, Emergence of scaling in random networks, Science 286 (1999) 509-512.
[6] S.H.Strogatz, Exploring complex networks, Nature 401(8) (2001) 268–276.
[7] D. Meunier, R. Lambiotte, A. Fornito, K.D. Ersche, E.T. Bullmore, Hierarchical modularity in human brain functional networks, Frontiers in Neuroinformatics 3 (37) (2009) 1-12.
[8] D. Meunier, R. Lambiotte, E.T. Bullmore, Modular and hierarchically modular organization of brain networks, Frontiers in Neuroscience 4 (200) (2010) 1-11.
[9] V.D. Blondel, J.L. Guillaume, R. Lambiotte, E. Lefebvre, Fast unfolding of communities in large networks, J. Stat. Mech. Theory E, 10 (2008) P10008.
[10] E. Ravasz, A.L. Somera, D.A. Mongru, Z.N. Oltvai, A.-L. Barabási, Hierarchical organization of modularity in metabolic networks, Science 297(2002) 1551-1555.
[11] E. Ravasz, A.-L. Barab ́asi, Hierarchical organization in complex networks, Physical Review E 67(2003) 026112.
[12] M.B. Eisen, P.T. Spellman, P.O. Brown, D. Botstein, Cluster analysis and display of genome-wideexpression patterns, Proc Natl Acad Sci USA 95 (1998) 14863.
[13] Clauset, C. M.E.J. Moore, Newman, Hierarchical structure and the prediction of missing links in networks, Nature 453 (7191) (2008) 98-101.
[14] C.M. Song, S. Havlin, A. Makse, Self-similarity of complex networks, Nature 433 (2005) 392-395.
[15] M. Girvan, M.E.J. Newman, Community structure in social and biological networks, Proc Natl Acad Sci USA 99 (2002) 7821-7826.
[16] M.E.J. Newman, Analysis of weighted networks, Physical Review E 70 (5) (2004) 05613.
[17] J. Scott, Soicial Network Analysis: A Handbook 2nd ed ,Sage Publications Inc, 2000.
[18] M.E.J. Newman, Fast algorithm for detecting community structure in networks, Physical Review E 69 (2004) 066133.
[19] J.M.Yang, An Application of Simon's Theory on the Architecture of Complex Systems, IEEE transactions on systems, man, and cybernetics, 23 (1) (1993) 264-267.
[20] R. Sitte, About the Predictability and Complexity of Complex Systems, M.A. Aziz-Alaoui, C. Bertelle, eds, From System Complexity to Emergent Properties, Springer, Berlin/ Heidelberg (2009) 23-48.
[21] W. Leontief, Input-Output Economics, Oxford Univ. Press, Oxford, UK, 1986.
[22] R. Tarjan, Depth-First Search and Linear Graph Algorithms, SIAM Journal on Computing 1 (2) (1972) 146–160.
[23] H. B. Gu, Analysis and evaluation of Input-Output price change model, The Journal of Quantitative & Technical Economics 6 (1994) 50-54.

**Fig. 1.** Various schematics of the room building

(a) (b) (c) (d)

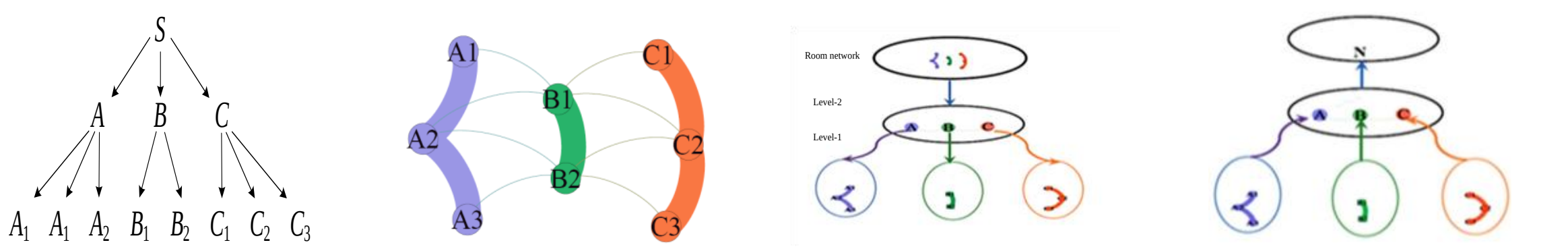

**(a)** The NDH of a room system under heat exchange interactions: S denotes the room system. A, B and C denote large rooms and the rest of the letters denote small rooms. The arrows point to the lower subsystem. **(b)** The network model of the room building: The vertices represent small rooms, an edge represents there is heat diffusion between two small rooms, and the thickness of the edge indicates the intensity of the heat diffusion. **(c)** The NDH of the room network (not showing weights): It is composed of one level-2 subnetwork and the level-2 subnetwork can be decomposed into three level-1 subnetworks. **(d)** Generation process of the room network.

**Fig. 2.** The process of analyzing the room network NDH

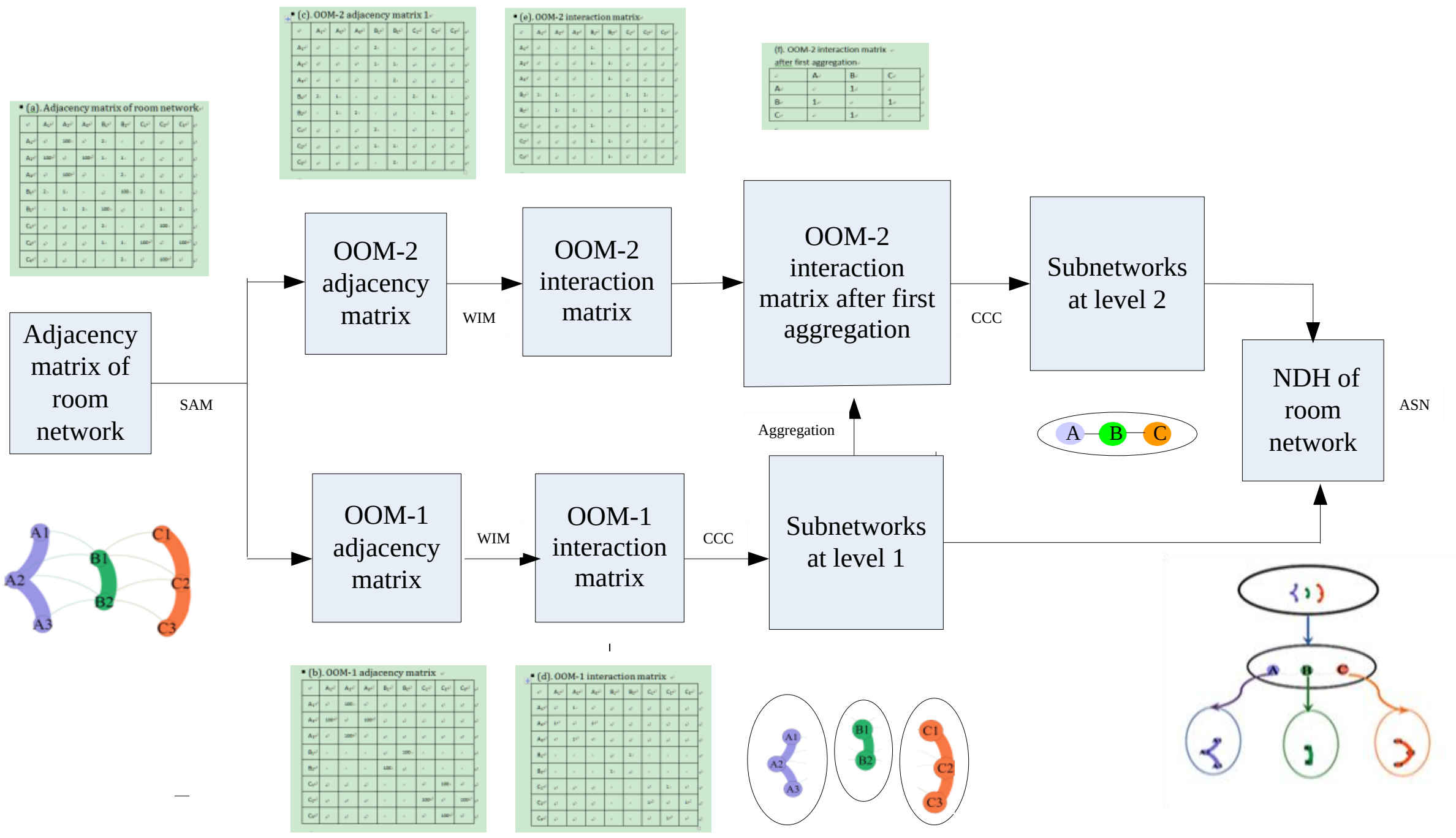

Note: SAM: Splitting the adjacency matrix. WIM: Writing the interaction matrices. CCC: calculating the connected components. ASN: Arranging subnetworks.

**Fig. 4.** GDIO network and the CPI triggered by Industry 1 in this network

(a) (b) (c) (d) (e) (f)

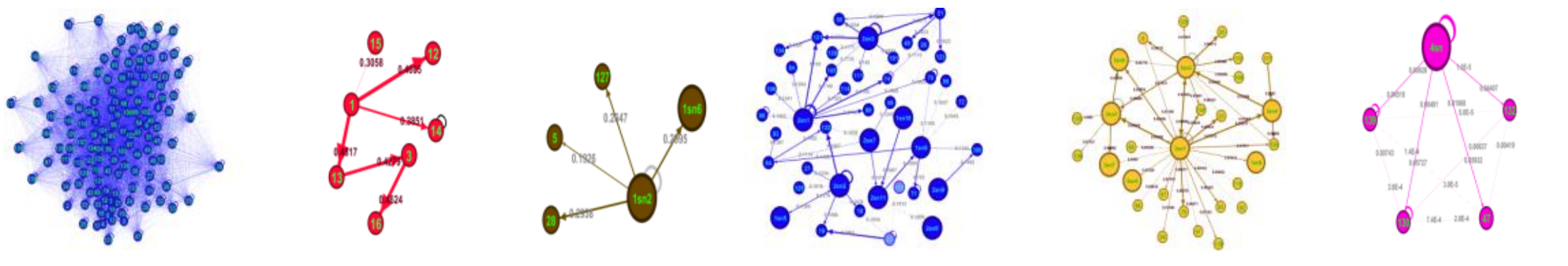

**(a)** GDIO network: The vertices in the Figure are represented by blue circles, and the green numbers in the circles are the sequence numbers of the industries. The edges are represented by blue lines and the edge thicknesses indicate the edge weights. **(b)** The CPI in level-1 subnetwork 1sn2. **(c)** The CPI in level-2 subnetwork 2sn2. **(d)** The CPI in level-3 subnetwork 3sn1. **(e)** The CPI in level-4 subnetwork 4sn. **(f)** The CPI in level-5 subnetwork 5sn. For clarity, the loop weights are not shown in the Figure. The relevant loop weights are: 0.3467 for vertex 14 in a; 0.2839 for AV 1sn2 in b; 0.1841 for AV 2sn2 in c.

**Fig. 3.** The NDH (not showing directions and weights) of GDIO network

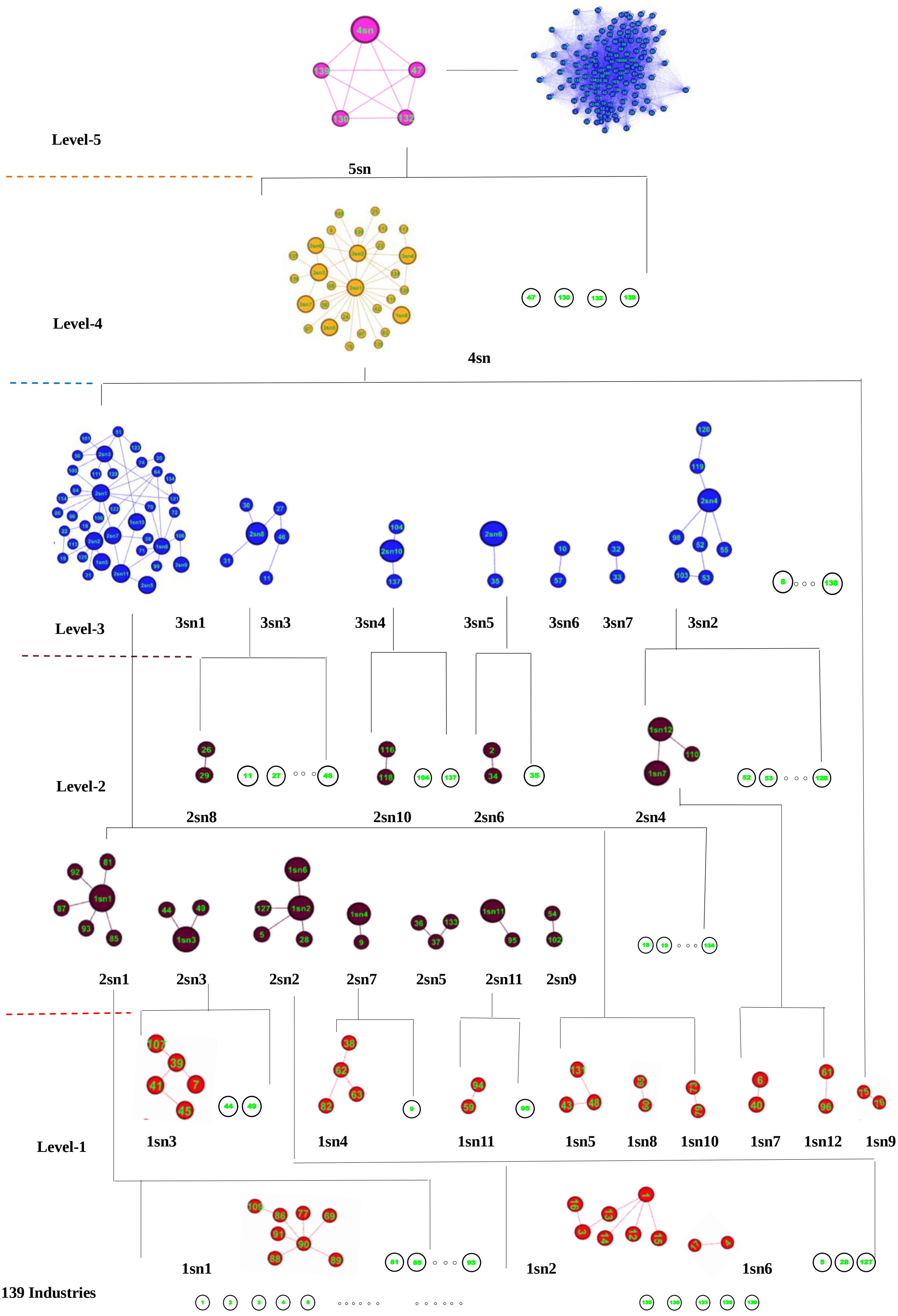

## Appendix I. The NDH of GD industry system

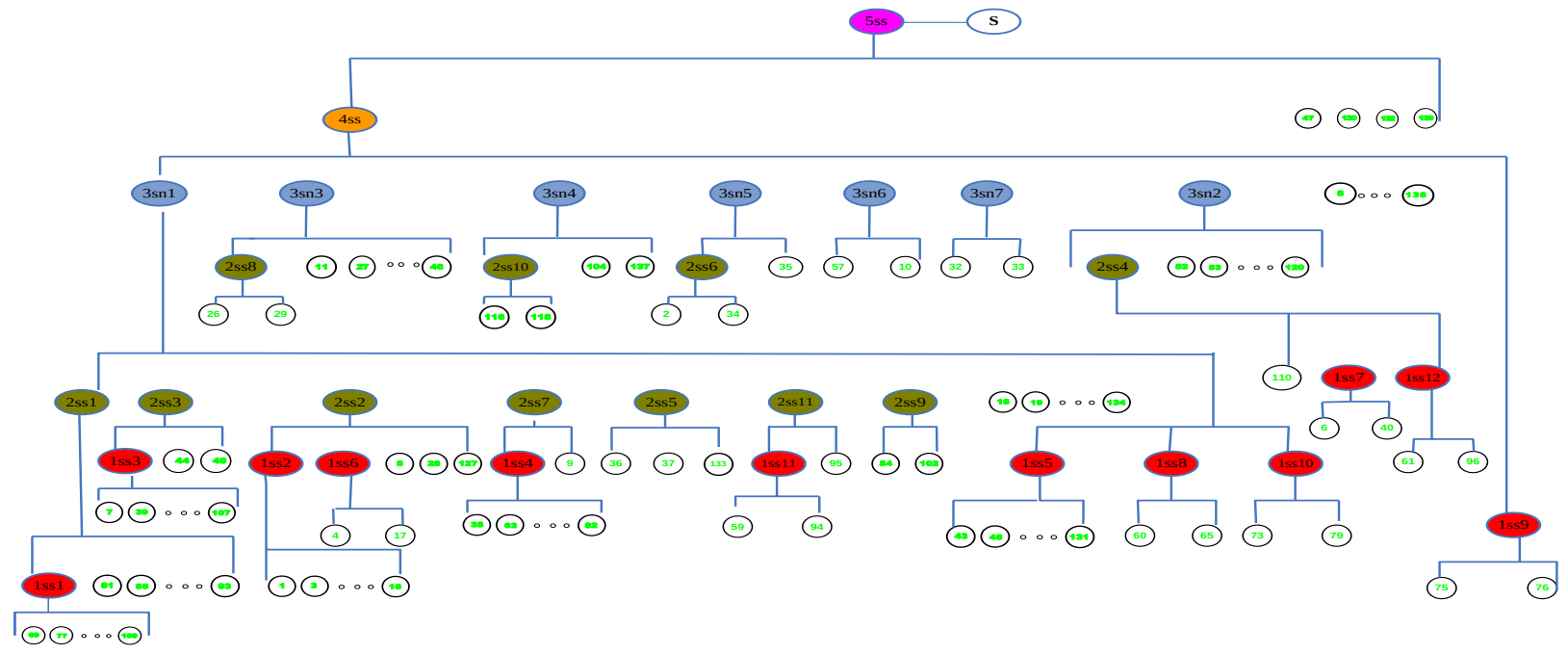

## Appendix II: Block diagonal form of several OOM-1 matrices of GDIO network

| | 107 | 39 | 7 | 41 | 45 | 38 | 62 | 63 | 82 | 59 | 94 | 131 | 43 | 48 | 60 | 65 | 109 | 86 | 77 | 69 | 91 | 90 | 89 | 88 | 16 | 3 | 13 | 1 | 14 | 12 | 15 | 17 | 4 | 40 | 6 | 61 | 96 | 75 | 76 | 79 | 73 |
|---|---|---|---|---|---|---|---|---|---|---|---|---|---|---|---|---|---|---|---|---|---|---|---|---|---|---|---|---|---|---|---|---|---|---|---|---|---|---|---|---|---|
| 107 | 0 | 1 | 0 | 0 | 0 | 0 | 0 | 0 | 0 | 0 | 0 | 0 | 0 | 0 | 0 | 0 | 0 | 0 | 0 | 0 | 0 | 0 | 0 | 0 | 0 | 0 | 0 | 0 | 0 | 0 | 0 | 0 | 0 | 0 | 0 | 0 | 0 | 0 | 0 | 0 | 0 |
| 39 | 1 | 0 | 1 | 1 | 0 | 0 | 0 | 0 | 0 | 0 | 0 | 0 | 0 | 0 | 0 | 0 | 0 | 0 | 0 | 0 | 0 | 0 | 0 | 0 | 0 | 0 | 0 | 0 | 0 | 0 | 0 | 0 | 0 | 0 | 0 | 0 | 0 | 0 | 0 | 0 | 0 |
| 7 | 0 | 1 | 0 | 0 | 0 | 0 | 0 | 0 | 0 | 0 | 0 | 0 | 0 | 0 | 0 | 0 | 0 | 0 | 0 | 0 | 0 | 0 | 0 | 0 | 0 | 0 | 0 | 0 | 0 | 0 | 0 | 0 | 0 | 0 | 0 | 0 | 0 | 0 | 0 | 0 | 0 |
| 41 | 0 | 1 | 0 | 0 | 1 | 0 | 0 | 0 | 0 | 0 | 0 | 0 | 0 | 0 | 0 | 0 | 0 | 0 | 0 | 0 | 0 | 0 | 0 | 0 | 0 | 0 | 0 | 0 | 0 | 0 | 0 | 0 | 0 | 0 | 0 | 0 | 0 | 0 | 0 | 0 | 0 |
| 45 | 0 | 0 | 0 | 1 | 0 | 0 | 0 | 0 | 0 | 0 | 0 | 0 | 0 | 0 | 0 | 0 | 0 | 0 | 0 | 0 | 0 | 0 | 0 | 0 | 0 | 0 | 0 | 0 | 0 | 0 | 0 | 0 | 0 | 0 | 0 | 0 | 0 | 0 | 0 | 0 | 0 |
| 38 | 0 | 0 | 0 | 0 | 0 | 0 | 1 | 0 | 0 | 0 | 0 | 0 | 0 | 0 | 0 | 0 | 0 | 0 | 0 | 0 | 0 | 0 | 0 | 0 | 0 | 0 | 0 | 0 | 0 | 0 | 0 | 0 | 0 | 0 | 0 | 0 | 0 | 0 | 0 | 0 | 0 |
| 62 | 0 | 0 | 0 | 0 | 0 | 1 | 0 | 1 | 1 | 0 | 0 | 0 | 0 | 0 | 0 | 0 | 0 | 0 | 0 | 0 | 0 | 0 | 0 | 0 | 0 | 0 | 0 | 0 | 0 | 0 | 0 | 0 | 0 | 0 | 0 | 0 | 0 | 0 | 0 | 0 | 0 |
| 63 | 0 | 0 | 0 | 0 | 0 | 0 | 1 | 0 | 0 | 0 | 0 | 0 | 0 | 0 | 0 | 0 | 0 | 0 | 0 | 0 | 0 | 0 | 0 | 0 | 0 | 0 | 0 | 0 | 0 | 0 | 0 | 0 | 0 | 0 | 0 | 0 | 0 | 0 | 0 | 0 | 0 |
| 82 | 0 | 0 | 0 | 0 | 0 | 0 | 1 | 0 | 0 | 0 | 0 | 0 | 0 | 0 | 0 | 0 | 0 | 0 | 0 | 0 | 0 | 0 | 0 | 0 | 0 | 0 | 0 | 0 | 0 | 0 | 0 | 0 | 0 | 0 | 0 | 0 | 0 | 0 | 0 | 0 | 0 |
| 59 | 0 | 0 | 0 | 0 | 0 | 0 | 0 | 0 | 0 | 0 | 1 | 0 | 0 | 0 | 0 | 0 | 0 | 0 | 0 | 0 | 0 | 0 | 0 | 0 | 0 | 0 | 0 | 0 | 0 | 0 | 0 | 0 | 0 | 0 | 0 | 0 | 0 | 0 | 0 | 0 | 0 |
| 94 | 0 | 0 | 0 | 0 | 0 | 0 | 0 | 0 | 0 | 1 | 0 | 0 | 0 | 0 | 0 | 0 | 0 | 0 | 0 | 0 | 0 | 0 | 0 | 0 | 0 | 0 | 0 | 0 | 0 | 0 | 0 | 0 | 0 | 0 | 0 | 0 | 0 | 0 | 0 | 0 | 0 |
| 131 | 0 | 0 | 0 | 0 | 0 | 0 | 0 | 0 | 0 | 0 | 0 | 0 | 0 | 1 | 0 | 0 | 0 | 0 | 0 | 0 | 0 | 0 | 0 | 0 | 0 | 0 | 0 | 0 | 0 | 0 | 0 | 0 | 0 | 0 | 0 | 0 | 0 | 0 | 0 | 0 | 0 |
| 43 | 0 | 0 | 0 | 0 | 0 | 0 | 0 | 0 | 0 | 0 | 0 | 0 | 0 | 1 | 0 | 0 | 0 | 0 | 0 | 0 | 0 | 0 | 0 | 0 | 0 | 0 | 0 | 0 | 0 | 0 | 0 | 0 | 0 | 0 | 0 | 0 | 0 | 0 | 0 | 0 | 0 |
| 48 | 0 | 0 | 0 | 0 | 0 | 0 | 0 | 0 | 0 | 0 | 0 | 1 | 1 | 0 | 0 | 0 | 0 | 0 | 0 | 0 | 0 | 0 | 0 | 0 | 0 | 0 | 0 | 0 | 0 | 0 | 0 | 0 | 0 | 0 | 0 | 0 | 0 | 0 | 0 | 0 | 0 |
| 60 | 0 | 0 | 0 | 0 | 0 | 0 | 0 | 0 | 0 | 0 | 0 | 0 | 0 | 0 | 0 | 1 | 0 | 0 | 0 | 0 | 0 | 0 | 0 | 0 | 0 | 0 | 0 | 0 | 0 | 0 | 0 | 0 | 0 | 0 | 0 | 0 | 0 | 0 | 0 | 0 | 0 |
| 65 | 0 | 0 | 0 | 0 | 0 | 0 | 0 | 0 | 0 | 0 | 0 | 0 | 0 | 0 | 1 | 0 | 0 | 0 | 0 | 0 | 0 | 0 | 0 | 0 | 0 | 0 | 0 | 0 | 0 | 0 | 0 | 0 | 0 | 0 | 0 | 0 | 0 | 0 | 0 | 0 | 0 |
| 109 | 0 | 0 | 0 | 0 | 0 | 0 | 0 | 0 | 0 | 0 | 0 | 0 | 0 | 0 | 0 | 0 | 0 | 1 | 0 | 0 | 0 | 0 | 0 | 0 | 0 | 0 | 0 | 0 | 0 | 0 | 0 | 0 | 0 | 0 | 0 | 0 | 0 | 0 | 0 | 0 | 0 |
| 86 | 0 | 0 | 0 | 0 | 0 | 0 | 0 | 0 | 0 | 0 | 0 | 0 | 0 | 0 | 0 | 0 | 1 | 0 | 0 | 0 | 0 | 1 | 0 | 0 | 0 | 0 | 0 | 0 | 0 | 0 | 0 | 0 | 0 | 0 | 0 | 0 | 0 | 0 | 0 | 0 | 0 |
| 77 | 0 | 0 | 0 | 0 | 0 | 0 | 0 | 0 | 0 | 0 | 0 | 0 | 0 | 0 | 0 | 0 | 0 | 0 | 0 | 0 | 0 | 1 | 0 | 0 | 0 | 0 | 0 | 0 | 0 | 0 | 0 | 0 | 0 | 0 | 0 | 0 | 0 | 0 | 0 | 0 | 0 |
| 69 | 0 | 0 | 0 | 0 | 0 | 0 | 0 | 0 | 0 | 0 | 0 | 0 | 0 | 0 | 0 | 0 | 0 | 0 | 0 | 0 | 0 | 1 | 0 | 0 | 0 | 0 | 0 | 0 | 0 | 0 | 0 | 0 | 0 | 0 | 0 | 0 | 0 | 0 | 0 | 0 | 0 |
| 91 | 0 | 0 | 0 | 0 | 0 | 0 | 0 | 0 | 0 | 0 | 0 | 0 | 0 | 0 | 0 | 0 | 0 | 0 | 0 | 0 | 0 | 1 | 0 | 0 | 0 | 0 | 0 | 0 | 0 | 0 | 0 | 0 | 0 | 0 | 0 | 0 | 0 | 0 | 0 | 0 | 0 |
| 90 | 0 | 0 | 0 | 0 | 0 | 0 | 0 | 0 | 0 | 0 | 0 | 0 | 0 | 0 | 0 | 0 | 0 | 1 | 1 | 1 | 1 | 0 | 1 | 1 | 0 | 0 | 0 | 0 | 0 | 0 | 0 | 0 | 0 | 0 | 0 | 0 | 0 | 0 | 0 | 0 | 0 |
| 89 | 0 | 0 | 0 | 0 | 0 | 0 | 0 | 0 | 0 | 0 | 0 | 0 | 0 | 0 | 0 | 0 | 0 | 0 | 0 | 0 | 0 | 1 | 0 | 0 | 0 | 0 | 0 | 0 | 0 | 0 | 0 | 0 | 0 | 0 | 0 | 0 | 0 | 0 | 0 | 0 | 0 |
| 88 | 0 | 0 | 0 | 0 | 0 | 0 | 0 | 0 | 0 | 0 | 0 | 0 | 0 | 0 | 0 | 0 | 0 | 0 | 0 | 0 | 0 | 1 | 0 | 0 | 0 | 0 | 0 | 0 | 0 | 0 | 0 | 0 | 0 | 0 | 0 | 0 | 0 | 0 | 0 | 0 | 0 |
| 16 | 0 | 0 | 0 | 0 | 0 | 0 | 0 | 0 | 0 | 0 | 0 | 0 | 0 | 0 | 0 | 0 | 0 | 0 | 0 | 0 | 0 | 0 | 0 | 0 | 0 | 1 | 0 | 0 | 0 | 0 | 0 | 0 | 0 | 0 | 0 | 0 | 0 | 0 | 0 | 0 | 0 |
| 3 | 0 | 0 | 0 | 0 | 0 | 0 | 0 | 0 | 0 | 0 | 0 | 0 | 0 | 0 | 0 | 0 | 0 | 0 | 0 | 0 | 0 | 0 | 0 | 0 | 1 | 0 | 1 | 0 | 0 | 0 | 0 | 0 | 0 | 0 | 0 | 0 | 0 | 0 | 0 | 0 | 0 |
| 13 | 0 | 0 | 0 | 0 | 0 | 0 | 0 | 0 | 0 | 0 | 0 | 0 | 0 | 0 | 0 | 0 | 0 | 0 | 0 | 0 | 0 | 0 | 0 | 0 | 0 | 1 | 0 | 1 | 0 | 0 | 0 | 0 | 0 | 0 | 0 | 0 | 0 | 0 | 0 | 0 | 0 |
| 1 | 0 | 0 | 0 | 0 | 0 | 0 | 0 | 0 | 0 | 0 | 0 | 0 | 0 | 0 | 0 | 0 | 0 | 0 | 0 | 0 | 0 | 0 | 0 | 0 | 0 | 0 | 1 | 0 | 1 | 1 | 1 | 0 | 0 | 0 | 0 | 0 | 0 | 0 | 0 | 0 | 0 |
| 14 | 0 | 0 | 0 | 0 | 0 | 0 | 0 | 0 | 0 | 0 | 0 | 0 | 0 | 0 | 0 | 0 | 0 | 0 | 0 | 0 | 0 | 0 | 0 | 0 | 0 | 0 | 0 | 1 | 0 | 0 | 0 | 0 | 0 | 0 | 0 | 0 | 0 | 0 | 0 | 0 | 0 |
| 12 | 0 | 0 | 0 | 0 | 0 | 0 | 0 | 0 | 0 | 0 | 0 | 0 | 0 | 0 | 0 | 0 | 0 | 0 | 0 | 0 | 0 | 0 | 0 | 0 | 0 | 0 | 0 | 1 | 0 | 0 | 0 | 0 | 0 | 0 | 0 | 0 | 0 | 0 | 0 | 0 | 0 |
| 15 | 0 | 0 | 0 | 0 | 0 | 0 | 0 | 0 | 0 | 0 | 0 | 0 | 0 | 0 | 0 | 0 | 0 | 0 | 0 | 0 | 0 | 0 | 0 | 0 | 0 | 0 | 0 | 1 | 0 | 0 | 0 | 0 | 0 | 0 | 0 | 0 | 0 | 0 | 0 | 0 | 0 |
| 17 | 0 | 0 | 0 | 0 | 0 | 0 | 0 | 0 | 0 | 0 | 0 | 0 | 0 | 0 | 0 | 0 | 0 | 0 | 0 | 0 | 0 | 0 | 0 | 0 | 0 | 0 | 0 | 0 | 0 | 0 | 0 | 0 | 1 | 0 | 0 | 0 | 0 | 0 | 0 | 0 | 0 |
| 4 | 0 | 0 | 0 | 0 | 0 | 0 | 0 | 0 | 0 | 0 | 0 | 0 | 0 | 0 | 0 | 0 | 0 | 0 | 0 | 0 | 0 | 0 | 0 | 0 | 0 | 0 | 0 | 0 | 0 | 0 | 0 | 1 | 0 | 0 | 0 | 0 | 0 | 0 | 0 | 0 | 0 |
| 40 | 0 | 0 | 0 | 0 | 0 | 0 | 0 | 0 | 0 | 0 | 0 | 0 | 0 | 0 | 0 | 0 | 0 | 0 | 0 | 0 | 0 | 0 | 0 | 0 | 0 | 0 | 0 | 0 | 0 | 0 | 0 | 0 | 0 | 0 | 1 | 0 | 0 | 0 | 0 | 0 | 0 |
| 6 | 0 | 0 | 0 | 0 | 0 | 0 | 0 | 0 | 0 | 0 | 0 | 0 | 0 | 0 | 0 | 0 | 0 | 0 | 0 | 0 | 0 | 0 | 0 | 0 | 0 | 0 | 0 | 0 | 0 | 0 | 0 | 0 | 0 | 1 | 0 | 0 | 0 | 0 | 0 | 0 | 0 |
| 61 | 0 | 0 | 0 | 0 | 0 | 0 | 0 | 0 | 0 | 0 | 0 | 0 | 0 | 0 | 0 | 0 | 0 | 0 | 0 | 0 | 0 | 0 | 0 | 0 | 0 | 0 | 0 | 0 | 0 | 0 | 0 | 0 | 0 | 0 | 0 | 0 | 1 | 0 | 0 | 0 | 0 |
| 96 | 0 | 0 | 0 | 0 | 0 | 0 | 0 | 0 | 0 | 0 | 0 | 0 | 0 | 0 | 0 | 0 | 0 | 0 | 0 | 0 | 0 | 0 | 0 | 0 | 0 | 0 | 0 | 0 | 0 | 0 | 0 | 0 | 0 | 0 | 0 | 1 | 0 | 0 | 0 | 0 | 0 |
| 75 | 0 | 0 | 0 | 0 | 0 | 0 | 0 | 0 | 0 | 0 | 0 | 0 | 0 | 0 | 0 | 0 | 0 | 0 | 0 | 0 | 0 | 0 | 0 | 0 | 0 | 0 | 0 | 0 | 0 | 0 | 0 | 0 | 0 | 0 | 0 | 0 | 0 | 0 | 1 | 0 | 0 |
| 76 | 0 | 0 | 0 | 0 | 0 | 0 | 0 | 0 | 0 | 0 | 0 | 0 | 0 | 0 | 0 | 0 | 0 | 0 | 0 | 0 | 0 | 0 | 0 | 0 | 0 | 0 | 0 | 0 | 0 | 0 | 0 | 0 | 0 | 0 | 0 | 0 | 0 | 1 | 0 | 0 | 0 |
| 79 | 0 | 0 | 0 | 0 | 0 | 0 | 0 | 0 | 0 | 0 | 0 | 0 | 0 | 0 | 0 | 0 | 0 | 0 | 0 | 0 | 0 | 0 | 0 | 0 | 0 | 0 | 0 | 0 | 0 | 0 | 0 | 0 | 0 | 0 | 0 | 0 | 0 | 0 | 0 | 0 | 1 |
| 73 | 0 | 0 | 0 | 0 | 0 | 0 | 0 | 0 | 0 | 0 | 0 | 0 | 0 | 0 | 0 | 0 | 0 | 0 | 0 | 0 | 0 | 0 | 0 | 0 | 0 | 0 | 0 | 0 | 0 | 0 | 0 | 0 | 0 | 0 | 0 | 0 | 0 | 0 | 0 | 1 | 0 |

**Fig. (1)**: Block diagonal form of the symmetric matrix of OOM-1 interaction matrix of GDIO network, a 0-1, symmetric matrix with all 0 diagonal elements. It just shows the non-zero part of the 139*139 matrix, the first row and column are the sequence number of the industry vertices (the same below).

| | 107 | 39 | 7 | 41 | 45 | 38 | 62 | 63 | 82 | 59 | 94 | 131 | 43 | 48 | 60 | 65 | 109 | 86 | 77 | 69 | 91 | 90 | 89 | 88 | 16 | 3 | 13 | 1 | 14 | 12 | 15 | 17 | 4 | 40 | 6 | 61 | 96 | 75 | 76 | 79 | 73 |
|---|---|---|---|---|---|---|---|---|---|---|---|---|---|---|---|---|---|---|---|---|---|---|---|---|---|---|---|---|---|---|---|---|---|---|---|---|---|---|---|---|---|
| 107 | 0 | 0 | 0 | 0 | 0 | 0 | 0 | 0 | 0 | 0 | 0 | 0 | 0 | 0 | 0 | 0 | 0 | 0 | 0 | 0 | 0 | 0 | 0 | 0 | 0 | 0 | 0 | 0 | 0 | 0 | 0 | 0 | 0 | 0 | 0 | 0 | 0 | 0 | 0 | 0 | 0 |
| 39 | 1 | 0 | 0 | 1 | 0 | 0 | 0 | 0 | 0 | 0 | 0 | 0 | 0 | 0 | 0 | 0 | 0 | 0 | 0 | 0 | 0 | 0 | 0 | 0 | 0 | 0 | 0 | 0 | 0 | 0 | 0 | 0 | 0 | 0 | 0 | 0 | 0 | 0 | 0 | 0 | 0 |
| 7 | 0 | 1 | 0 | 0 | 0 | 0 | 0 | 0 | 0 | 0 | 0 | 0 | 0 | 0 | 0 | 0 | 0 | 0 | 0 | 0 | 0 | 0 | 0 | 0 | 0 | 0 | 0 | 0 | 0 | 0 | 0 | 0 | 0 | 0 | 0 | 0 | 0 | 0 | 0 | 0 | 0 |
| 41 | 0 | 0 | 0 | 0 | 1 | 0 | 0 | 0 | 0 | 0 | 0 | 0 | 0 | 0 | 0 | 0 | 0 | 0 | 0 | 0 | 0 | 0 | 0 | 0 | 0 | 0 | 0 | 0 | 0 | 0 | 0 | 0 | 0 | 0 | 0 | 0 | 0 | 0 | 0 | 0 | 0 |
| 45 | 0 | 0 | 0 | 0 | 1 | 0 | 0 | 0 | 0 | 0 | 0 | 0 | 0 | 0 | 0 | 0 | 0 | 0 | 0 | 0 | 0 | 0 | 0 | 0 | 0 | 0 | 0 | 0 | 0 | 0 | 0 | 0 | 0 | 0 | 0 | 0 | 0 | 0 | 0 | 0 | 0 |
| 38 | 0 | 0 | 0 | 0 | 0 | 0 | 0 | 0 | 0 | 0 | 0 | 0 | 0 | 0 | 0 | 0 | 0 | 0 | 0 | 0 | 0 | 0 | 0 | 0 | 0 | 0 | 0 | 0 | 0 | 0 | 0 | 0 | 0 | 0 | 0 | 0 | 0 | 0 | 0 | 0 | 0 |
| 62 | 0 | 0 | 0 | 0 | 0 | 1 | 1 | 1 | 1 | 0 | 0 | 0 | 0 | 0 | 0 | 0 | 0 | 0 | 0 | 0 | 0 | 0 | 0 | 0 | 0 | 0 | 0 | 0 | 0 | 0 | 0 | 0 | 0 | 0 | 0 | 0 | 0 | 0 | 0 | 0 | 0 |
| 63 | 0 | 0 | 0 | 0 | 0 | 0 | 0 | 0 | 0 | 0 | 0 | 0 | 0 | 0 | 0 | 0 | 0 | 0 | 0 | 0 | 0 | 0 | 0 | 0 | 0 | 0 | 0 | 0 | 0 | 0 | 0 | 0 | 0 | 0 | 0 | 0 | 0 | 0 | 0 | 0 | 0 |
| 82 | 0 | 0 | 0 | 0 | 0 | 0 | 0 | 0 | 0 | 0 | 0 | 0 | 0 | 0 | 0 | 0 | 0 | 0 | 0 | 0 | 0 | 0 | 0 | 0 | 0 | 0 | 0 | 0 | 0 | 0 | 0 | 0 | 0 | 0 | 0 | 0 | 0 | 0 | 0 | 0 | 0 |
| 59 | 0 | 0 | 0 | 0 | 0 | 0 | 0 | 0 | 0 | 0 | 0 | 0 | 0 | 0 | 0 | 0 | 0 | 0 | 0 | 0 | 0 | 0 | 0 | 0 | 0 | 0 | 0 | 0 | 0 | 0 | 0 | 0 | 0 | 0 | 0 | 0 | 0 | 0 | 0 | 0 | 0 |
| 94 | 0 | 0 | 0 | 0 | 0 | 0 | 0 | 0 | 0 | 1 | 1 | 0 | 0 | 0 | 0 | 0 | 0 | 0 | 0 | 0 | 0 | 0 | 0 | 0 | 0 | 0 | 0 | 0 | 0 | 0 | 0 | 0 | 0 | 0 | 0 | 0 | 0 | 0 | 0 | 0 | 0 |
| 131 | 0 | 0 | 0 | 0 | 0 | 0 | 0 | 0 | 0 | 0 | 0 | 0 | 0 | 0 | 0 | 0 | 0 | 0 | 0 | 0 | 0 | 0 | 0 | 0 | 0 | 0 | 0 | 0 | 0 | 0 | 0 | 0 | 0 | 0 | 0 | 0 | 0 | 0 | 0 | 0 | 0 |
| 43 | 0 | 0 | 0 | 0 | 0 | 0 | 0 | 0 | 0 | 0 | 0 | 0 | 0 | 0 | 0 | 0 | 0 | 0 | 0 | 0 | 0 | 0 | 0 | 0 | 0 | 0 | 0 | 0 | 0 | 0 | 0 | 0 | 0 | 0 | 0 | 0 | 0 | 0 | 0 | 0 | 0 |
| 48 | 0 | 0 | 0 | 0 | 0 | 0 | 0 | 0 | 0 | 0 | 0 | 1 | 1 | 0 | 0 | 0 | 0 | 0 | 0 | 0 | 0 | 0 | 0 | 0 | 0 | 0 | 0 | 0 | 0 | 0 | 0 | 0 | 0 | 0 | 0 | 0 | 0 | 0 | 0 | 0 | 0 |
| 60 | 0 | 0 | 0 | 0 | 0 | 0 | 0 | 0 | 0 | 0 | 0 | 0 | 0 | 0 | 0 | 1 | 0 | 0 | 0 | 0 | 0 | 0 | 0 | 0 | 0 | 0 | 0 | 0 | 0 | 0 | 0 | 0 | 0 | 0 | 0 | 0 | 0 | 0 | 0 | 0 | 0 |
| 65 | 0 | 0 | 0 | 0 | 0 | 0 | 0 | 0 | 0 | 0 | 0 | 0 | 0 | 0 | 0 | 0 | 0 | 0 | 0 | 0 | 0 | 0 | 0 | 0 | 0 | 0 | 0 | 0 | 0 | 0 | 0 | 0 | 0 | 0 | 0 | 0 | 0 | 0 | 0 | 0 | 0 |
| 109 | 0 | 0 | 0 | 0 | 0 | 0 | 0 | 0 | 0 | 0 | 0 | 0 | 0 | 0 | 0 | 0 | 0 | 0 | 0 | 0 | 0 | 0 | 0 | 0 | 0 | 0 | 0 | 0 | 0 | 0 | 0 | 0 | 0 | 0 | 0 | 0 | 0 | 0 | 0 | 0 | 0 |
| 86 | 0 | 0 | 0 | 0 | 0 | 0 | 0 | 0 | 0 | 0 | 0 | 0 | 0 | 0 | 0 | 0 | 1 | 0 | 0 | 0 | 0 | 0 | 0 | 0 | 0 | 0 | 0 | 0 | 0 | 0 | 0 | 0 | 0 | 0 | 0 | 0 | 0 | 0 | 0 | 0 | 0 |
| 77 | 0 | 0 | 0 | 0 | 0 | 0 | 0 | 0 | 0 | 0 | 0 | 0 | 0 | 0 | 0 | 0 | 0 | 0 | 0 | 0 | 0 | 0 | 0 | 0 | 0 | 0 | 0 | 0 | 0 | 0 | 0 | 0 | 0 | 0 | 0 | 0 | 0 | 0 | 0 | 0 | 0 |
| 69 | 0 | 0 | 0 | 0 | 0 | 0 | 0 | 0 | 0 | 0 | 0 | 0 | 0 | 0 | 0 | 0 | 0 | 0 | 0 | 0 | 0 | 0 | 0 | 0 | 0 | 0 | 0 | 0 | 0 | 0 | 0 | 0 | 0 | 0 | 0 | 0 | 0 | 0 | 0 | 0 | 0 |
| 91 | 0 | 0 | 0 | 0 | 0 | 0 | 0 | 0 | 0 | 0 | 0 | 0 | 0 | 0 | 0 | 0 | 0 | 0 | 0 | 0 | 1 | 0 | 0 | 0 | 0 | 0 | 0 | 0 | 0 | 0 | 0 | 0 | 0 | 0 | 0 | 0 | 0 | 0 | 0 | 0 | 0 |
| 90 | 0 | 0 | 0 | 0 | 0 | 0 | 0 | 0 | 0 | 0 | 0 | 0 | 0 | 0 | 0 | 0 | 0 | 1 | 1 | 1 | 1 | 1 | 1 | 1 | 0 | 0 | 0 | 0 | 0 | 0 | 0 | 0 | 0 | 0 | 0 | 0 | 0 | 0 | 0 | 0 | 0 |
| 89 | 0 | 0 | 0 | 0 | 0 | 0 | 0 | 0 | 0 | 0 | 0 | 0 | 0 | 0 | 0 | 0 | 0 | 0 | 0 | 0 | 0 | 0 | 0 | 0 | 0 | 0 | 0 | 0 | 0 | 0 | 0 | 0 | 0 | 0 | 0 | 0 | 0 | 0 | 0 | 0 | 0 |
| 88 | 0 | 0 | 0 | 0 | 0 | 0 | 0 | 0 | 0 | 0 | 0 | 0 | 0 | 0 | 0 | 0 | 0 | 0 | 0 | 0 | 0 | 0 | 0 | 0 | 0 | 0 | 0 | 0 | 0 | 0 | 0 | 0 | 0 | 0 | 0 | 0 | 0 | 0 | 0 | 0 | 0 |
| 16 | 0 | 0 | 0 | 0 | 0 | 0 | 0 | 0 | 0 | 0 | 0 | 0 | 0 | 0 | 0 | 0 | 0 | 0 | 0 | 0 | 0 | 0 | 0 | 0 | 0 | 0 | 0 | 0 | 0 | 0 | 0 | 0 | 0 | 0 | 0 | 0 | 0 | 0 | 0 | 0 | 0 |
| 3 | 0 | 0 | 0 | 0 | 0 | 0 | 0 | 0 | 0 | 0 | 0 | 0 | 0 | 0 | 0 | 0 | 0 | 0 | 0 | 0 | 0 | 0 | 0 | 0 | 1 | 0 | 0 | 0 | 0 | 0 | 0 | 0 | 0 | 0 | 0 | 0 | 0 | 0 | 0 | 0 | 0 |
| 13 | 0 | 0 | 0 | 0 | 0 | 0 | 0 | 0 | 0 | 0 | 0 | 0 | 0 | 0 | 0 | 0 | 0 | 0 | 0 | 0 | 0 | 0 | 0 | 0 | 0 | 1 | 0 | 0 | 0 | 0 | 0 | 0 | 0 | 0 | 0 | 0 | 0 | 0 | 0 | 0 | 0 |
| 1 | 0 | 0 | 0 | 0 | 0 | 0 | 0 | 0 | 0 | 0 | 0 | 0 | 0 | 0 | 0 | 0 | 0 | 0 | 0 | 0 | 0 | 0 | 0 | 0 | 0 | 0 | 1 | 0 | 1 | 1 | 1 | 0 | 0 | 0 | 0 | 0 | 0 | 0 | 0 | 0 | 0 |
| 14 | 0 | 0 | 0 | 0 | 0 | 0 | 0 | 0 | 0 | 0 | 0 | 0 | 0 | 0 | 0 | 0 | 0 | 0 | 0 | 0 | 0 | 0 | 0 | 0 | 0 | 0 | 0 | 0 | 1 | 0 | 0 | 0 | 0 | 0 | 0 | 0 | 0 | 0 | 0 | 0 | 0 |
| 12 | 0 | 0 | 0 | 0 | 0 | 0 | 0 | 0 | 0 | 0 | 0 | 0 | 0 | 0 | 0 | 0 | 0 | 0 | 0 | 0 | 0 | 0 | 0 | 0 | 0 | 0 | 0 | 0 | 0 | 0 | 0 | 0 | 0 | 0 | 0 | 0 | 0 | 0 | 0 | 0 | 0 |
| 15 | 0 | 0 | 0 | 0 | 0 | 0 | 0 | 0 | 0 | 0 | 0 | 0 | 0 | 0 | 0 | 0 | 0 | 0 | 0 | 0 | 0 | 0 | 0 | 0 | 0 | 0 | 0 | 0 | 0 | 0 | 0 | 0 | 0 | 0 | 0 | 0 | 0 | 0 | 0 | 0 | 0 |
| 17 | 0 | 0 | 0 | 0 | 0 | 0 | 0 | 0 | 0 | 0 | 0 | 0 | 0 | 0 | 0 | 0 | 0 | 0 | 0 | 0 | 0 | 0 | 0 | 0 | 0 | 0 | 0 | 0 | 0 | 0 | 0 | 0 | 0 | 0 | 0 | 0 | 0 | 0 | 0 | 0 | 0 |
| 4 | 0 | 0 | 0 | 0 | 0 | 0 | 0 | 0 | 0 | 0 | 0 | 0 | 0 | 0 | 0 | 0 | 0 | 0 | 0 | 0 | 0 | 0 | 0 | 0 | 0 | 0 | 0 | 0 | 0 | 0 | 0 | 1 | 0 | 0 | 0 | 0 | 0 | 0 | 0 | 0 | 0 |
| 40 | 0 | 0 | 0 | 0 | 0 | 0 | 0 | 0 | 0 | 0 | 0 | 0 | 0 | 0 | 0 | 0 | 0 | 0 | 0 | 0 | 0 | 0 | 0 | 0 | 0 | 0 | 0 | 0 | 0 | 0 | 0 | 0 | 0 | 0 | 0 | 0 | 0 | 0 | 0 | 0 | 0 |
| 6 | 0 | 0 | 0 | 0 | 0 | 0 | 0 | 0 | 0 | 0 | 0 | 0 | 0 | 0 | 0 | 0 | 0 | 0 | 0 | 0 | 0 | 0 | 0 | 0 | 0 | 0 | 0 | 0 | 0 | 0 | 0 | 0 | 0 | 1 | 0 | 0 | 0 | 0 | 0 | 0 | 0 |
| 61 | 0 | 0 | 0 | 0 | 0 | 0 | 0 | 0 | 0 | 0 | 0 | 0 | 0 | 0 | 0 | 0 | 0 | 0 | 0 | 0 | 0 | 0 | 0 | 0 | 0 | 0 | 0 | 0 | 0 | 0 | 0 | 0 | 0 | 0 | 0 | 0 | 0 | 0 | 0 | 0 | 0 |
| 96 | 0 | 0 | 0 | 0 | 0 | 0 | 0 | 0 | 0 | 0 | 0 | 0 | 0 | 0 | 0 | 0 | 0 | 0 | 0 | 0 | 0 | 0 | 0 | 0 | 0 | 0 | 0 | 0 | 0 | 0 | 0 | 0 | 0 | 0 | 0 | 1 | 1 | 0 | 0 | 0 | 0 |
| 75 | 0 | 0 | 0 | 0 | 0 | 0 | 0 | 0 | 0 | 0 | 0 | 0 | 0 | 0 | 0 | 0 | 0 | 0 | 0 | 0 | 0 | 0 | 0 | 0 | 0 | 0 | 0 | 0 | 0 | 0 | 0 | 0 | 0 | 0 | 0 | 0 | 0 | 0 | 0 | 0 | 0 |
| 76 | 0 | 0 | 0 | 0 | 0 | 0 | 0 | 0 | 0 | 0 | 0 | 0 | 0 | 0 | 0 | 0 | 0 | 0 | 0 | 0 | 0 | 0 | 0 | 0 | 0 | 0 | 0 | 0 | 0 | 0 | 0 | 0 | 0 | 0 | 0 | 0 | 0 | 1 | 1 | 0 | 0 |
| 79 | 0 | 0 | 0 | 0 | 0 | 0 | 0 | 0 | 0 | 0 | 0 | 0 | 0 | 0 | 0 | 0 | 0 | 0 | 0 | 0 | 0 | 0 | 0 | 0 | 0 | 0 | 0 | 0 | 0 | 0 | 0 | 0 | 0 | 0 | 0 | 0 | 0 | 0 | 0 | 1 | 1 |
| 73 | 0 | 0 | 0 | 0 | 0 | 0 | 0 | 0 | 0 | 0 | 0 | 0 | 0 | 0 | 0 | 0 | 0 | 0 | 0 | 0 | 0 | 0 | 0 | 0 | 0 | 0 | 0 | 0 | 0 | 0 | 0 | 0 | 0 | 0 | 0 | 0 | 0 | 0 | 0 | 0 | 0 |

**Fig. (2):** Block diagonal form of OOM-1 interaction matrix of GDIO network, a 0-1, non-symmetric matrix with not all 0 diagonal elements.

| | 107 | 39 | 7 | 41 | 45 | 38 | 62 | 63 | 82 | 59 | 94 | 131 | 43 | 48 | 60 | 65 | 109 | 86 | 77 | 69 | 91 | 90 | 89 | 88 | 16 | 3 | 13 | 1 | 14 | 12 | 15 | 17 | 4 | 40 | 6 | 61 | 96 | 75 | 76 | 79 | 73 |
|---|---|---|---|---|---|---|---|---|---|---|---|---|---|---|---|---|---|---|---|---|---|---|---|---|---|---|---|---|---|---|---|---|---|---|---|---|---|---|---|---|---|
| 107 | 0 | 0 | 0 | 0 | 0 | 0 | 0 | 0 | 0 | 0 | 0 | 0 | 0 | 0 | 0 | 0 | 0 | 0 | 0 | 0 | 0 | 0 | 0 | 0 | 0 | 0 | 0 | 0 | 0 | 0 | 0 | 0 | 0 | 0 | 0 | 0 | 0 | 0 | 0 | 0 | 0 |
| 39 | 0.3464 | 0 | 0 | 0.3587 | 0 | 0 | 0 | 0 | 0 | 0 | 0 | 0 | 0 | 0 | 0 | 0 | 0 | 0 | 0 | 0 | 0 | 0 | 0 | 0 | 0 | 0 | 0 | 0 | 0 | 0 | 0 | 0 | 0 | 0 | 0 | 0 | 0 | 0 | 0 | 0 | 0 |
| 7 | 0 | 0.4463 | 0 | 0 | 0 | 0 | 0 | 0 | 0 | 0 | 0 | 0 | 0 | 0 | 0 | 0 | 0 | 0 | 0 | 0 | 0 | 0 | 0 | 0 | 0 | 0 | 0 | 0 | 0 | 0 | 0 | 0 | 0 | 0 | 0 | 0 | 0 | 0 | 0 | 0 | 0 |
| 41 | 0 | 0 | 0 | 0 | 0.3056 | 0 | 0 | 0 | 0 | 0 | 0 | 0 | 0 | 0 | 0 | 0 | 0 | 0 | 0 | 0 | 0 | 0 | 0 | 0 | 0 | 0 | 0 | 0 | 0 | 0 | 0 | 0 | 0 | 0 | 0 | 0 | 0 | 0 | 0 | 0 | 0 |
| 45 | 0 | 0 | 0 | 0 | 0 | 0 | 0 | 0 | 0 | 0 | 0 | 0 | 0 | 0 | 0 | 0 | 0 | 0 | 0 | 0 | 0 | 0 | 0 | 0 | 0 | 0 | 0 | 0 | 0 | 0 | 0 | 0 | 0 | 0 | 0 | 0 | 0 | 0 | 0 | 0 | 0 |
| 38 | 0 | 0 | 0 | 0 | 0 | 0 | 0 | 0 | 0 | 0 | 0 | 0 | 0 | 0 | 0 | 0 | 0 | 0 | 0 | 0 | 0 | 0 | 0 | 0 | 0 | 0 | 0 | 0 | 0 | 0 | 0 | 0 | 0 | 0 | 0 | 0 | 0 | 0 | 0 | 0 | 0 |
| 62 | 0 | 0 | 0 | 0 | 0 | 0.3346 | 0 | 0.4367 | 0.3482 | 0 | 0 | 0 | 0 | 0 | 0 | 0 | 0 | 0 | 0 | 0 | 0 | 0 | 0 | 0 | 0 | 0 | 0 | 0 | 0 | 0 | 0 | 0 | 0 | 0 | 0 | 0 | 0 | 0 | 0 | 0 | 0 |
| 63 | 0 | 0 | 0 | 0 | 0 | 0 | 0 | 0 | 0 | 0 | 0 | 0 | 0 | 0 | 0 | 0 | 0 | 0 | 0 | 0 | 0 | 0 | 0 | 0 | 0 | 0 | 0 | 0 | 0 | 0 | 0 | 0 | 0 | 0 | 0 | 0 | 0 | 0 | 0 | 0 | 0 |
| 82 | 0 | 0 | 0 | 0 | 0 | 0 | 0 | 0 | 0 | 0 | 0 | 0 | 0 | 0 | 0 | 0 | 0 | 0 | 0 | 0 | 0 | 0 | 0 | 0 | 0 | 0 | 0 | 0 | 0 | 0 | 0 | 0 | 0 | 0 | 0 | 0 | 0 | 0 | 0 | 0 | 0 |
| 59 | 0 | 0 | 0 | 0 | 0 | 0 | 0 | 0 | 0 | 0 | 0 | 0 | 0 | 0 | 0 | 0 | 0 | 0 | 0 | 0 | 0 | 0 | 0 | 0 | 0 | 0 | 0 | 0 | 0 | 0 | 0 | 0 | 0 | 0 | 0 | 0 | 0 | 0 | 0 | 0 | 0 |
| 94 | 0 | 0 | 0 | 0 | 0 | 0 | 0 | 0 | 0 | 0.5269 | 0 | 0 | 0 | 0 | 0 | 0 | 0 | 0 | 0 | 0 | 0 | 0 | 0 | 0 | 0 | 0 | 0 | 0 | 0 | 0 | 0 | 0 | 0 | 0 | 0 | 0 | 0 | 0 | 0 | 0 | 0 |
| 131 | 0 | 0 | 0 | 0 | 0 | 0 | 0 | 0 | 0 | 0 | 0 | 0 | 0 | 0 | 0 | 0 | 0 | 0 | 0 | 0 | 0 | 0 | 0 | 0 | 0 | 0 | 0 | 0 | 0 | 0 | 0 | 0 | 0 | 0 | 0 | 0 | 0 | 0 | 0 | 0 | 0 |
| 43 | 0 | 0 | 0 | 0 | 0 | 0 | 0 | 0 | 0 | 0 | 0 | 0 | 0 | 0 | 0 | 0 | 0 | 0 | 0 | 0 | 0 | 0 | 0 | 0 | 0 | 0 | 0 | 0 | 0 | 0 | 0 | 0 | 0 | 0 | 0 | 0 | 0 | 0 | 0 | 0 | 0 |
| 48 | 0 | 0 | 0 | 0 | 0 | 0 | 0 | 0 | 0 | 0 | 0 | 0.3104 | 0.5682 | 0 | 0 | 0 | 0 | 0 | 0 | 0 | 0 | 0 | 0 | 0 | 0 | 0 | 0 | 0 | 0 | 0 | 0 | 0 | 0 | 0 | 0 | 0 | 0 | 0 | 0 | 0 | 0 |
| 60 | 0 | 0 | 0 | 0 | 0 | 0 | 0 | 0 | 0 | 0 | 0 | 0 | 0 | 0 | 0 | 0.4451 | 0 | 0 | 0 | 0 | 0 | 0 | 0 | 0 | 0 | 0 | 0 | 0 | 0 | 0 | 0 | 0 | 0 | 0 | 0 | 0 | 0 | 0 | 0 | 0 | 0 |
| 65 | 0 | 0 | 0 | 0 | 0 | 0 | 0 | 0 | 0 | 0 | 0 | 0 | 0 | 0 | 0 | 0 | 0 | 0 | 0 | 0 | 0 | 0 | 0 | 0 | 0 | 0 | 0 | 0 | 0 | 0 | 0 | 0 | 0 | 0 | 0 | 0 | 0 | 0 | 0 | 0 | 0 |
| 109 | 0 | 0 | 0 | 0 | 0 | 0 | 0 | 0 | 0 | 0 | 0 | 0 | 0 | 0 | 0 | 0 | 0 | 0 | 0 | 0 | 0 | 0 | 0 | 0 | 0 | 0 | 0 | 0 | 0 | 0 | 0 | 0 | 0 | 0 | 0 | 0 | 0 | 0 | 0 | 0 | 0 |
| 86 | 0 | 0 | 0 | 0 | 0 | 0 | 0 | 0 | 0 | 0 | 0 | 0 | 0 | 0 | 0 | 0 | 0.6437 | 0 | 0 | 0 | 0 | 0 | 0 | 0 | 0 | 0 | 0 | 0 | 0 | 0 | 0 | 0 | 0 | 0 | 0 | 0 | 0 | 0 | 0 | 0 | 0 |
| 77 | 0 | 0 | 0 | 0 | 0 | 0 | 0 | 0 | 0 | 0 | 0 | 0 | 0 | 0 | 0 | 0 | 0 | 0 | 0 | 0 | 0 | 0 | 0 | 0 | 0 | 0 | 0 | 0 | 0 | 0 | 0 | 0 | 0 | 0 | 0 | 0 | 0 | 0 | 0 | 0 | 0 |
| 69 | 0 | 0 | 0 | 0 | 0 | 0 | 0 | 0 | 0 | 0 | 0 | 0 | 0 | 0 | 0 | 0 | 0 | 0 | 0 | 0 | 0 | 0 | 0 | 0 | 0 | 0 | 0 | 0 | 0 | 0 | 0 | 0 | 0 | 0 | 0 | 0 | 0 | 0 | 0 | 0 | 0 |
| 91 | 0 | 0 | 0 | 0 | 0 | 0 | 0 | 0 | 0 | 0 | 0 | 0 | 0 | 0 | 0 | 0 | 0 | 0 | 0 | 0 | 0 | 0 | 0 | 0 | 0 | 0 | 0 | 0 | 0 | 0 | 0 | 0 | 0 | 0 | 0 | 0 | 0 | 0 | 0 | 0 | 0 |
| 90 | 0 | 0 | 0 | 0 | 0 | 0 | 0 | 0 | 0 | 0 | 0 | 0 | 0 | 0 | 0 | 0 | 0 | 0.4374 | 0.4902 | 0.4673 | 0.3484 | 0 | 0.5101 | 0.4935 | 0 | 0 | 0 | 0 | 0 | 0 | 0 | 0 | 0 | 0 | 0 | 0 | 0 | 0 | 0 | 0 | 0 |
| 89 | 0 | 0 | 0 | 0 | 0 | 0 | 0 | 0 | 0 | 0 | 0 | 0 | 0 | 0 | 0 | 0 | 0 | 0 | 0 | 0 | 0 | 0 | 0 | 0 | 0 | 0 | 0 | 0 | 0 | 0 | 0 | 0 | 0 | 0 | 0 | 0 | 0 | 0 | 0 | 0 | 0 |
| 88 | 0 | 0 | 0 | 0 | 0 | 0 | 0 | 0 | 0 | 0 | 0 | 0 | 0 | 0 | 0 | 0 | 0 | 0 | 0 | 0 | 0 | 0 | 0 | 0 | 0 | 0 | 0 | 0 | 0 | 0 | 0 | 0 | 0 | 0 | 0 | 0 | 0 | 0 | 0 | 0 | 0 |
| 16 | 0 | 0 | 0 | 0 | 0 | 0 | 0 | 0 | 0 | 0 | 0 | 0 | 0 | 0 | 0 | 0 | 0 | 0 | 0 | 0 | 0 | 0 | 0 | 0 | 0 | 0 | 0 | 0 | 0 | 0 | 0 | 0 | 0 | 0 | 0 | 0 | 0 | 0 | 0 | 0 | 0 |
| 3 | 0 | 0 | 0 | 0 | 0 | 0 | 0 | 0 | 0 | 0 | 0 | 0 | 0 | 0 | 0 | 0 | 0 | 0 | 0 | 0 | 0 | 0 | 0 | 0 | 0.4624 | 0 | 0 | 0 | 0 | 0 | 0 | 0 | 0 | 0 | 0 | 0 | 0 | 0 | 0 | 0 | 0 |
| 13 | 0 | 0 | 0 | 0 | 0 | 0 | 0 | 0 | 0 | 0 | 0 | 0 | 0 | 0 | 0 | 0 | 0 | 0 | 0 | 0 | 0 | 0 | 0 | 0 | 0 | 0.4279 | 0 | 0 | 0 | 0 | 0 | 0 | 0 | 0 | 0 | 0 | 0 | 0 | 0 | 0 | 0 |
| 1 | 0 | 0 | 0 | 0 | 0 | 0 | 0 | 0 | 0 | 0 | 0 | 0 | 0 | 0 | 0 | 0 | 0 | 0 | 0 | 0 | 0 | 0 | 0 | 0 | 0 | 0 | 0.4817 | 0 | 0.3851 | 0.4605 | 0.3058 | 0 | 0 | 0 | 0 | 0 | 0 | 0 | 0 | 0 | 0 |
| 14 | 0 | 0 | 0 | 0 | 0 | 0 | 0 | 0 | 0 | 0 | 0 | 0 | 0 | 0 | 0 | 0 | 0 | 0 | 0 | 0 | 0 | 0 | 0 | 0 | 0 | 0 | 0 | 0 | 0 | 0 | 0 | 0 | 0 | 0 | 0 | 0 | 0 | 0 | 0 | 0 | 0 |
| 12 | 0 | 0 | 0 | 0 | 0 | 0 | 0 | 0 | 0 | 0 | 0 | 0 | 0 | 0 | 0 | 0 | 0 | 0 | 0 | 0 | 0 | 0 | 0 | 0 | 0 | 0 | 0 | 0 | 0 | 0 | 0 | 0 | 0 | 0 | 0 | 0 | 0 | 0 | 0 | 0 | 0 |
| 15 | 0 | 0 | 0 | 0 | 0 | 0 | 0 | 0 | 0 | 0 | 0 | 0 | 0 | 0 | 0 | 0 | 0 | 0 | 0 | 0 | 0 | 0 | 0 | 0 | 0 | 0 | 0 | 0 | 0 | 0 | 0 | 0 | 0 | 0 | 0 | 0 | 0 | 0 | 0 | 0 | 0 |
| 17 | 0 | 0 | 0 | 0 | 0 | 0 | 0 | 0 | 0 | 0 | 0 | 0 | 0 | 0 | 0 | 0 | 0 | 0 | 0 | 0 | 0 | 0 | 0 | 0 | 0 | 0 | 0 | 0 | 0 | 0 | 0 | 0 | 0 | 0 | 0 | 0 | 0 | 0 | 0 | 0 | 0 |
| 4 | 0 | 0 | 0 | 0 | 0 | 0 | 0 | 0 | 0 | 0 | 0 | 0 | 0 | 0 | 0 | 0 | 0 | 0 | 0 | 0 | 0 | 0 | 0 | 0 | 0 | 0 | 0 | 0 | 0 | 0 | 0 | 0.503 | 0 | 0 | 0 | 0 | 0 | 0 | 0 | 0 | 0 |
| 40 | 0 | 0 | 0 | 0 | 0 | 0 | 0 | 0 | 0 | 0 | 0 | 0 | 0 | 0 | 0 | 0 | 0 | 0 | 0 | 0 | 0 | 0 | 0 | 0 | 0 | 0 | 0 | 0 | 0 | 0 | 0 | 0 | 0 | 0 | 0 | 0 | 0 | 0 | 0 | 0 | 0 |
| 6 | 0 | 0 | 0 | 0 | 0 | 0 | 0 | 0 | 0 | 0 | 0 | 0 | 0 | 0 | 0 | 0 | 0 | 0 | 0 | 0 | 0 | 0 | 0 | 0 | 0 | 0 | 0 | 0 | 0 | 0 | 0 | 0 | 0 | 0.8 | 0 | 0 | 0 | 0 | 0 | 0 | 0 |
| 61 | 0 | 0 | 0 | 0 | 0 | 0 | 0 | 0 | 0 | 0 | 0 | 0 | 0 | 0 | 0 | 0 | 0 | 0 | 0 | 0 | 0 | 0 | 0 | 0 | 0 | 0 | 0 | 0 | 0 | 0 | 0 | 0 | 0 | 0 | 0 | 0 | 0 | 0 | 0 | 0 | 0 |
| 96 | 0 | 0 | 0 | 0 | 0 | 0 | 0 | 0 | 0 | 0 | 0 | 0 | 0 | 0 | 0 | 0 | 0 | 0 | 0 | 0 | 0 | 0 | 0 | 0 | 0 | 0 | 0 | 0 | 0 | 0 | 0 | 0 | 0 | 0 | 0 | 0.631 | 0 | 0 | 0 | 0 | 0 |
| 75 | 0 | 0 | 0 | 0 | 0 | 0 | 0 | 0 | 0 | 0 | 0 | 0 | 0 | 0 | 0 | 0 | 0 | 0 | 0 | 0 | 0 | 0 | 0 | 0 | 0 | 0 | 0 | 0 | 0 | 0 | 0 | 0 | 0 | 0 | 0 | 0 | 0 | 0 | 0 | 0 | 0 |
| 76 | 0 | 0 | 0 | 0 | 0 | 0 | 0 | 0 | 0 | 0 | 0 | 0 | 0 | 0 | 0 | 0 | 0 | 0 | 0 | 0 | 0 | 0 | 0 | 0 | 0 | 0 | 0 | 0 | 0 | 0 | 0 | 0 | 0 | 0 | 0 | 0 | 0 | 0.4031 | 0 | 0 | 0 |
| 79 | 0 | 0 | 0 | 0 | 0 | 0 | 0 | 0 | 0 | 0 | 0 | 0 | 0 | 0 | 0 | 0 | 0 | 0 | 0 | 0 | 0 | 0 | 0 | 0 | 0 | 0 | 0 | 0 | 0 | 0 | 0 | 0 | 0 | 0 | 0 | 0 | 0 | 0 | 0 | 0 | 0.3889 |
| 73 | 0 | 0 | 0 | 0 | 0 | 0 | 0 | 0 | 0 | 0 | 0 | 0 | 0 | 0 | 0 | 0 | 0 | 0 | 0 | 0 | 0 | 0 | 0 | 0 | 0 | 0 | 0 | 0 | 0 | 0 | 0 | 0 | 0 | 0 | 0 | 0 | 0 | 0 | 0 | 0 | 0 |

**Fig. (3)**: Block diagonal form of the matrix 1 of GDIO network, a non-0-1, non-symmetric matrix with not all 0 diagonal elements.

---

[i] The first version of this paper was submitted to journals such as Nature Communications beginning in May 2021, and this is a revised version ready for submission to other journals.